%% file: main.tex
\documentclass[a4paper]{interact}

\newif\ifblinded
\blindedfalse   

\usepackage{epstopdf}
\usepackage[onehalfspacing]{setspace}
\theoremstyle{plain}

\theoremstyle{definition}

\theoremstyle{remark}

\graphicspath{{figures/}}
\usepackage{orcidlink}
\usepackage{subfiles}
\usepackage{float}
\usepackage{bbm}
\newcommand{\tr}{\operatorname{tr}}
\usepackage{caption,subcaption}

\RequirePackage[dvipsnames,svgnames,x11names,table]{xcolor}

\usepackage[natbib,style=apa,backend=biber]{biblatex}
\title{Bias-Reduced Estimation of Structural Equation Models}

\ifblinded
\else
\author{%
    \name{
        Haziq Jamil$\textsuperscript{1,2}$~\orcidlink{0000-0003-3298-1010}, 
        Yves Rosseel$\textsuperscript{3}$~\orcidlink{0000-0002-4129-4477}, 
        Oliver Kemp$\textsuperscript{4}$, 
        Ioannis Kosmidis$\textsuperscript{4}$~\orcidlink{0000-0003-1556-0302}}
    \affil{
        \textsuperscript{1} Computer, Electrical and Mathematical Sciences and Engineering (CEMSE) Division, King Abdullah University of Science and Technology, Thuwal, Kingdom of Saudi Arabia\\
        \textsuperscript{2} Mathematical Sciences, Faculty of Science, Universiti Brunei Darussalam, Bandar Seri Begawan, Brunei\\
        \textsuperscript{3} Department of Data Analysis, Ghent University, Ghent, Belgium\\
        \textsuperscript{4} Department of Statistics, University of Warwick, Coventry, United Kingdom
    }
}
\thanks{CONTACT Haziq Jamil. Email: haziq.jamil@kaust.edu.sa}
\fi

\begin{document}


\maketitle

\begin{abstract}
Finite-sample bias is a pervasive challenge in the estimation of structural equation models (SEMs), especially when sample sizes are small or measurement reliability is low. 
A range of methods have been proposed to improve finite-sample bias in the SEM literature, ranging from analytic bias corrections to resampling-based techniques, with each carrying trade-offs in scope, computational burden, and statistical performance. 
We introduce the reduced-bias M-estimation framework \parencite[RBM,][J. R. Stat. Soc. Series B Stat. Methodol.]{kosmidis2024empirical} to SEMs. 
The RBM framework is attractive as it requires only first- and second-order derivatives of the log-likelihood, which renders it both straightforward to implement, and computationally more efficient compared to resampling-based alternatives such as bootstrap and jackknife.
It is also robust to departures from modelling assumptions. 
Using the same simulation setup as in \textcite{dhaene2022resampling}, we illustrate that RBM estimators consistently reduce mean bias in the estimation of SEMs without inflating mean squared error.
They also deliver improvements in both median bias and inference relative to maximum likelihood estimators, while maintaining robustness under non-normality. 
Our findings suggest that RBM offers a promising, practical, and broadly applicable tool for mitigating bias in the estimation of SEMs, particularly in small-sample research contexts.
\end{abstract}


\section{Introduction}

Structural Equation Models (SEMs) are widely used statistical models that allow for the specification, estimation, and testing of complex relationships among observed and latent variables \parencite{Hoyle2023handbook, SkrondalBook2004}.
Despite their versatility and theoretical appeal, the practical use of SEMs in empirical research is often challenged by small sample sizes.
Small samples are particularly common in fields where data collection is expensive, time-consuming, or logistically constrained, such as clinical psychology, developmental research, and neuroscience \autocite{vandeschoot2020small}.
Small samples are also often an inevitable consequence of studying rare populations.
For example, research involving children with burned facial injuries is inherently limited by the small number of individuals who meet these specific criteria \autocite{quezada2016explanatory}.
Similarly, studies focusing on rare genetic disorders \autocite{figueroa2021structural}, elite athletes \autocite{hodge2009athlete,fabbricatore2023componentbased}, or unique cultural groups \autocite{manuela2013pacific} often face strict limitations in sample size due to the rarity of the target population.
In such contexts, the use of SEMs remains desirable for testing complex theoretical models, but the challenges posed by small samples become unavoidable and cannot be ignored.

When the sample size is small, SEMs face a range of well-documented difficulties, involving non-convergence of estimation algorithms, improper solutions such as negative variance estimates (Heywood cases) or non-positive definite estimates of variance-covariance matrices, inflated standard errors, and low power of statistical tests \autocite{deng2018structural,bentler1999structural,nevitt2004evaluating}.
In response to these challenges, researchers have developed a range of techniques to improve estimation performance in small samples.
These include alternative estimators such as those from Bayesian methods \autocite{muthen2012bayesian,lee2012basic}, penalized likelihood approaches \autocite{huang2017penalized,jacobucci2016regularized}, and model simplification strategies \autocite{rosseel2024structural}.
Such developments have substantially improved the feasibility and stability of SEMs in small-sample contexts.
However, even when convergence is achieved and estimates appear admissible, the estimators may exhibit poor finite-sample performance, which, in turn, can impact inferential conclusions.

Many attempts to improve finite-sample performance focus on reducing estimation bias.
For SEMs that are mathematically equivalent to mixed effects models \autocite{bauer2003estimating,cheung2013implementing,mcneish2016using,mcneish2018brief}, the use of restricted maximum likelihood (REML) has been found to result in less biased estimates of the variance components \autocite{corbeil1976restricted,patterson1971recovery}.
In a paper focusing on small sample corrections for Wald tests in SEMs, \textcite{ozenne2020small} describe an analytic method to correct for the asymptotic bias in the ML estimation of SEMs with exogenous variables.
Importantly, their method only targets the (observed or latent) (co)variance parameters in the model, and parameters like regression coefficients and factor loadings are not affected.
More general strategies towards reducing bias of the estimators for all parameters are the resampling-based approaches proposed by \textcite{dhaene2022resampling}, which adapt jackknife and bootstrap techniques to SEM.
Such methods are broadly applicable and relatively easy to implement.
However, they can be computationally intensive and exhibit limitations in very small samples.
These approaches represent important steps toward addressing finite-sample bias in SEM estimation,
but each carries trade-offs between generality, complexity, and computational cost.

Recently, \textcite{kosmidis2024empirical} introduced a new framework for reducing bias in M-estimators (including maximum likelihood estimators) derived from asymptotically unbiased estimating functions.
That framework operates either \emph{explicit}ly, by subtracting an estimate of the bias from the original estimator, or \emph{implicit}ly, by finding the roots of adjusted estimating functions.
Notably, reduced-bias M-estimation (RBM) requires only a bias approximation that involves the first and second derivatives of the estimating functions with respect to the parameters, eliminating the need for complex expectations under the model, and hence being more robust to model misspecification and easier to apply than other popular bias reduction methods, such as that of \textcite{firth1993bias}, which require expected values of products of log-likelihood derivatives with respect to the assumed model.
Explicit RBM (eRBM) estimation requires just the original estimates and an evaluation of the bias approximation, and the implicit (iRBM) version requires the computation of the root of adjusted score equations, with an adjustment that directly depends on that bias approximation.
Put simply, eRBM is a post hoc correction, while iRBM has built the correction into the estimation itself.
As a result, RBM estimation requires no resampling or repeated fitting of the SEM.
Furthermore, in contrast to existing bias-reduction techniques for SEMs, the RBM approach is fully general and, hence, applicable to any SEM specification, while simultaneously targeting all model parameters. 
In addition, the RBM estimators have the same asymptotic distribution as the original estimators, ensuring that standard inference and model selection procedures apply directly, including those that are robust to departures of the normality assumptions.

Motivated by these appealing properties, the goal of this paper is to translate the RBM technology 
to the SEM framework and to investigate its performance by extending the simulation setup
of \textcite{dhaene2022resampling}, who investigated resampling-based approaches to bias reduction for two types of SEMs: a two-factor SEM, and a latent growth curve model.
In addition, we provide accompanying R code in the supplementary materials to facilitate its application.

The remainder of the paper is structured as follows.
We begin by briefly outlining the SEM framework and introducing the notation used throughout the paper in section 2.
In section 3, we then present a concise overview of several bias-reduction techniques, including the method proposed by \textcite{ozenne2020small}, the jackknife and bootstrap approaches, and the recently introduced RBM method.
Following this, in sections 4 and 5, we report the results of two simulation studies designed to evaluate the performance of the RBM approach in comparison to a selection of alternatives.
Finally, in section 6, we conclude with a discussion of the findings and offer practical recommendations for applied researchers.

\section{Structural Equation Models}\label{structural-equation-models}

A structural equation model (SEM) is a multivariate modelling approach that combines aspects of regression and factor analysis to simultaneously capture relationships involving both observed and latent variables.
It is typically described in terms of two components:
a measurement part and a structural part.
Let $\mathbf y_i\in\mathbb R^p$ be a vector of observed variables for the sample indexed by $i=1,\dots,n$.
The measurement part of the model is defined as
\begin{equation}
\mathbf y_i = \boldsymbol\nu + \boldsymbol\Lambda \boldsymbol\eta_i + \boldsymbol\epsilon_i, \label{eq-measurement}
\end{equation}
where \(\boldsymbol\nu\) is a \(p\)-vector of measurement intercepts, \(\boldsymbol\Lambda\) is a \(p
\times q\) matrix of factor loadings, \(\boldsymbol\eta_i\) is a \(q\)-vector of latent variables, and \(\boldsymbol\epsilon_i\) is a \(p\)-vector of measurement errors.
It is common and advantageous to assume that the vector $\boldsymbol\epsilon_i$ is a centred normal random vector with covariance matrix $\boldsymbol\Theta$.

Similarly, the vector $\boldsymbol\eta_i$ is also assumed to be normally distributed, as a consequence of the structural equations
\begin{equation}
\boldsymbol\eta_i = \boldsymbol\alpha + \mathbf B \boldsymbol \eta_i + \boldsymbol\zeta_i. \label{eq-structural}
\end{equation}
Here, \(\boldsymbol\alpha\) is a \(q\)-vector of latent intercepts, \(\mathbf B\) is a \(q \times q\) sparse matrix of latent regression coefficients, and \(\boldsymbol\zeta_i\) is a random \(q\)-vector of structural disturbances with mean zero and covariance matrix \(\boldsymbol\Psi\), assumed to be normally distributed.
To rule out circular dependencies, the matrix $\mathbf B$ is specified such that $\operatorname{diag}(\mathbf B) = \mathbf 0$ and $\mathbf I- \mathbf B$ is invertible \autocite{kaplan2009structural}.
We also assume for each $i\in\{1,\dots,n\}$ that $\operatorname{cov}(\epsilon_{ij},\eta_{ik}) = \operatorname{cov}(\epsilon_{ij}, \zeta_{ik}) = 0$, $j=1,\dots,p$ and $k=1,\dots,q$, ensuring that measurement errors are uncorrelated with the latent variables (and independent under normality).

Let $\vartheta$ denote the $m$-vector of all free and non-redundant parameters in the SEM, which includes the free entries of $\boldsymbol\nu$, $\boldsymbol\Lambda$, $\boldsymbol\Theta$, $\boldsymbol\alpha$, $\mathbf B$, and $\boldsymbol\Psi$.
For observed data that are continuous, these parameters are typically estimated by maximum likelihood (ML).
To ensure model identifiability, the number of free parameters must not exceed the number of distinct pieces of information available in the sample mean and covariance structures. 
To set the scale of the latent variables, additional constraints are required---for example, fixing one factor loading per latent variable for scale identification, or setting variance of latent variables to unity.
Additional constraints may be imposed to reflect substantive modelling intentions or to achieve a more parsimonious specification, as is common in growth curve models.

\subsection{Maximum Likelihood Estimation}

Equations \eqref{eq-measurement} and \eqref{eq-structural} together with the normality assumptions allow us to reduce the model to a $p$-variate normal distribution for the observed variables, i.e. $\mathbf y_i\sim \text{N}_p\big(\boldsymbol\mu(\vartheta), \boldsymbol\Sigma(\vartheta)\big)$ for $i=1,\dots,n$, where
\begin{align}
\boldsymbol\mu(\vartheta) &= \boldsymbol\nu + \boldsymbol\Lambda (\mathbf I - \mathbf B)^{-1} \boldsymbol\alpha, \label{eq-marg-mean} \\
\boldsymbol\Sigma(\vartheta) &= \boldsymbol\Sigma^*(\vartheta) + \boldsymbol\Theta \label{eq-marg-var},
\end{align}
with \(\boldsymbol\Sigma^*(\vartheta) = \boldsymbol\Lambda (\mathbf I - \mathbf B)^{-1} \boldsymbol\Psi (\mathbf I - \mathbf B)^{-\top} \boldsymbol\Lambda^\top\), and \(\mathbf M^{-\top} = (\mathbf M^{-1})^\top\) for a symmetric, positive-definite matrix $\mathbf M$.
Let $\bar{\mathbf y} = n^{-1} \sum_{i=1}^n \mathbf y_i$ and $\mathbf S = n^{-1} \sum_{i=1}^n (\mathbf y_i - \bar{\mathbf y})(\mathbf y_i - \bar{\mathbf y})^\top$ denote the sample mean and covariance matrix, respectively.
ML estimation proceeds by maximizing the log likelihood
\begin{equation}
\ell(\vartheta)
= -\frac{n}{2}\Bigl[
p \log(2\pi)
+ \log \bigl|\boldsymbol\Sigma(\vartheta)\bigr|
+ \operatorname{tr} \bigl(\boldsymbol\Sigma(\vartheta)^{-1} \mathbf S\bigr)
+ \bigl(\bar {\mathbf y} - \boldsymbol\mu(\vartheta)\bigr)^{\top}
  \boldsymbol\Sigma(\vartheta)^{-1}
  \bigl(\bar {\mathbf y} - \boldsymbol\mu(\vartheta)\bigr)
\Bigr]
\label{eq-sem_lik}
\end{equation}
with respect to \(\vartheta\) (or equivalently, by solving the first-order conditions $\nabla \ell(\vartheta)=\mathbf{0}$) to give the ML estimator \(\hat{\vartheta}\) of \(\vartheta\).
When no mean structure is specified in the SEM, the last term in \eqref{eq-sem_lik} vanishes, and the estimation reduces to a problem of fitting the model-implied covariance matrix $\Sigma(\vartheta)$ to the sample covariance matrix $\mathbf S$.

Denote by $\ell_i(\vartheta)$ the $i$th contribution to the log-likelihood, such that $\ell(\vartheta) = \sum_{i=1}^n \ell_i(\vartheta)$.
The ML estimator has the attractive property of being consistent and asymptotically normal, i.e. under standard regularity conditions  \parencite{cox1979theoretical} as $n\to\infty$,
\begin{equation}
\sqrt n (\hat\vartheta - \bar\vartheta) \xrightarrow{\;\;\text D\;\;} \text{N}_m\big(\mathbf 0, I(\bar\vartheta)^{-1} \big), \label{eq-asymptheta}
\end{equation}
where $\bar\vartheta$ is the true parameter vector, $I(\vartheta) = \mathbb{E}\left[ \nabla\ell_1(\vartheta)\nabla\ell_1(\vartheta)^\top \right]$ is the $m\times m$ Fisher information matrix for $\vartheta$ based on a single, generic observation (the subscript `$1$' is used purely for notational convenience), from which standard errors may be computed.
When model assumptions such as multivariate normality may be violated, standard errors can instead be based on a robust ``sandwich'' estimator of the asymptotic covariance matrix \autocite{savalei2022computational}
\begin{equation}
\sqrt n(\hat\vartheta - \bar\vartheta) \xrightarrow[]{\;\;\text D\;\;} \text N_m\left(\mathbf 0, \big[ U(\bar\vartheta)V(\bar\vartheta)^{-1} U(\bar\vartheta) \big]^{-1} \right), \label{eq-robustasymp}
\end{equation}
where $U(\vartheta) = \mathbb E \left[ -\nabla\nabla^\top \ell_1(\vartheta) \right]$ is the sensitivity matrix, and $V(\vartheta) = \operatorname{var} \left[ \nabla \ell_1(\vartheta)  \right]$ is the variability matrix.
For a correctly specified model, \eqref{eq-robustasymp} collapses to \eqref{eq-asymptheta} making use of the well-known Bartlett identities.
Robust or sandwich standard errors for SEM have been discussed in several works including \textcite{satorra1994corrections,savalei2014understanding}.


Despite the consistency property, ML estimators are generally biased in finite samples, where the bias is of asymptotic order $O(n^{-1})$ \autocite{cox1968general}.
This bias arises because the score function is not exactly centred at the true parameter value, causing the solution to the likelihood equations to be systematically shifted \autocite[see, e.g.,][]{firth1993bias,kosmidis2014bias}.
While mean parameters are exactly unbiased under normal-theory ML, other parameters---particularly variances, covariances, and coefficients that are nonlinear functions of them---may exhibit substantial finite-sample bias, which is often most pronounced in small samples.
These considerations motivate a variety of approaches aimed at reducing or correcting this bias, which are reviewed in the next section.

\section{Reducing Bias in SEM Estimation}\label{reducing-bias}

\subsection{Bias Reduction}\label{bias-reduction}

\textcite[Sec. 1]{kosmidis2024empirical} provide an integrated review of a range of general-purpose bias reduction procedures, elements of which we present here.
For an estimator $\hat\vartheta$ of $\vartheta$, let $\mathcal B(\bar\vartheta) = \mathbb E(\hat\vartheta - \bar\vartheta)$ be the bias function, with $\bar{\vartheta}$ denoting the true parameter value, as before.
The aim of bias reduction methods is to produce a new estimator $\vartheta^*$ of $\vartheta$ that approximates the solution of
\begin{equation}
\hat{\vartheta} - \vartheta^* = \mathcal B(\bar{\vartheta}) \label{eq-bias_eq}
\end{equation} 
with respect to \(\vartheta^*\).
However, the unbiased estimator \(\vartheta^*\) generally cannot be computed exactly since \(\mathcal B(\cdot)\) is typically not available in closed form and \(\bar{\vartheta}\) is not known.
Instead, the solution to (\ref{eq-bias_eq}) may be approximated to produce an estimator with asymptotically smaller bias than \(\hat{\vartheta}\).
Such approximations can be obtained either through analytical methods, using asymptotic bias corrections \autocite{cordeiro1991bias,efron1975defining} or adjusted score functions \autocite{firth1993bias,kosmidis2009bias}, or through simulation-based approaches such as the bootstrap and jackknife, discussed shortly.

Bias reduction methods may operate explicitly or implicitly.
Explicit methods aim to approximate the bias of an estimator, and then subtract this bias from the original estimator to produce a reduced-bias estimator.
In particular, such methods replace \(\mathcal B(\vartheta)\) in (\ref{eq-bias_eq}) with an estimate \(\mathcal B^*\), which is computed from the available data, before computing \(\vartheta^\dagger =
\hat{\vartheta} - \mathcal B^*\).
Implicit methods, on the other hand, compute a reduced-bias estimator by replacing \(\mathcal B(\vartheta)\) with \(\hat{\mathcal B}(\tilde{\vartheta})\), which is an estimator of the bias function at the desired estimator, and solve the implicit equation \(\hat{\vartheta} - \tilde{\vartheta} = \hat{\mathcal B}(\tilde{\vartheta})\) for \(\tilde{\vartheta}\).
Generally, explicit methods are simpler to compute and interpret (as being a ``bias-corrected'' version of the original estimate), while implicit methods tend to offer better performance as bias reduction is built into the estimation step itself. \vspace{-0.5em}

\subsection{Resampling-Based Methods}\label{sec-resampling}

Prominent examples of explicit bias reduction methods include the bootstrap \autocite{efron1994introduction,hall1988bootstrap} and jackknife \autocite{quenouille1956notes,efron1982jackknife}.
The bootstrap involves generating \(T\) bootstrap samples \(\{y^{*}_{(t)}\}_{t=1}^T\) by sampling with replacement from the original dataset.
Then, the model is fitted to each bootstrap sample to obtain estimates \(\hat\vartheta^{*}_{(1)}, \ldots, \hat\vartheta^{*}_{(T)}\).
The bootstrap bias-corrected estimator is then computed as
\[
\hat\vartheta_{\text{boot}} = 2\hat\vartheta - \frac{1}{T}\sum_{t=1}^T \hat\vartheta^{*}_{(t)}.
\]
The number of bootstrap replications $T$ is decided in advance and largely determines the total computational cost of the bootstrap procedure, although increasing $T$ generally improves the precision of the resulting estimate.
The jackknife on the other hand involves obtaining estimates \(\hat\vartheta_{(i)}\) by fitting the model to the \(n\) subsamples of size \(n - 1\) that result by leaving out one observation at a time.
This can be quick for small $n$ but becomes computationally demanding for larger datasets, as the model must be refitted $n$ times.
The jackknife bias-corrected estimator is then computed as
\[
\hat\vartheta_{\text{jack}} = n\hat\vartheta - (n-1)\frac{1}{n}\sum_{i=1}^n
\hat\vartheta_{(i)} \, . 
\]
Both approaches share the intuition of estimating the bias by repeatedly perturbing the data, measuring how the estimator responds, and then extrapolating back to what the estimate would have been with zero bias.

While intuitive and straightforward to implement, the bootstrap and jackknife each have their limitations.
In SEM estimation with small samples, convergence issues and estimator instability can lead to unreliable resampled estimates \autocite{dhaene2022resampling}, while in large samples, the procedures can be computationally intensive.
Additionally, the jackknife may underestimate variability for nonlinear estimators \autocite{efron1994introduction}, and the bootstrap can perform poorly when the sampling distribution is skewed or parameters are near the boundary of the parameter space \autocite[Sec. 5.2]{davison1997bootstrap}.

\subsection{Alternative Approaches to Reducing Bias}\label{sec-alternative}

Restricted maximum likelihood (REML), while formally not a bias reduction technique, is frequently used for correcting biases in the estimation of variance components \autocite{patterson1971recovery} in linear mixed effects \autocite{corbeil1976restricted}.
A range of SEMs can be expressed as a linear mixed effects model \(\mathbf y \mid \mathbf b \sim \text{N}(\mathbf X \boldsymbol\beta + \mathbf Z \mathbf b, \sigma^2 \mathbf I)\) \autocite{bauer2003estimating,cheung2013implementing}, where \(\mathbf X\) and \(\mathbf Z\) are the fixed and random model matrices, respectively, \(\boldsymbol\beta\) is the vector of fixed effects, and \(\mathbf b \sim N(\mathbf 0, \boldsymbol\Sigma_u)\) is the vector of random effects.
As a note, $\mathbf y$ here is understood in the ``long-format'' sense, differing from its earlier use in the SEM context.
REML estimates are obtained by maximizing a marginal likelihood function based on residuals from a least squares fit of the model \(\mathbf y \sim \text{N}(\mathbf X\boldsymbol\beta, \sigma^2 \mathbf I)\).
That marginal likelihood ends up not involving the fixed effects parameters.
Specifically, REML maximizes \autocite{fitzmaurice2011applied}
\[ 
\ell_{\text{\tiny REML}}(\vartheta) =
- \frac{1}{2} \log |\mathbf V(\vartheta)| - \frac{1}{2} \log | \mathbf X^\top \mathbf V(\vartheta)^{-1} \mathbf X | 
- \frac{1}{2} (\mathbf y - \mathbf X\hat{\boldsymbol\beta}_{\mathbf V(\vartheta)})^\top \mathbf V(\vartheta)^{-1} (\mathbf y - \mathbf X\hat{\boldsymbol\beta}_{\mathbf V(\vartheta)}) ,
\]
where \(\mathbf V(\vartheta) = \mathbf Z\boldsymbol\Sigma_u \mathbf Z^\top + \sigma^2 \mathbf I\) is the covariance matrix of the marginal distribution of \(\mathbf y\), and \(\hat{\boldsymbol\beta}_V = (\mathbf X^\top \mathbf V^{-1} \mathbf X)^{-1} \mathbf X^\top \mathbf V^{-1} \mathbf y\).
By integrating out the fixed effects, REML effectively adjusts for the loss of degrees of freedom associated with the estimation, leading to variance component estimates that are typically less biased than ML.

In a paper focusing on small sample corrections for Wald tests in SEMs with exogenous covariates, \textcite[Sec. 4]{ozenne2020small} derive the asymptotic bias of \(\hat{\boldsymbol\Sigma} = n^{-1} \sum_{i = 1}^n \big(\mathbf y_i - \boldsymbol\mu(\hat\vartheta)\big)^\top \big(\mathbf y_i - \boldsymbol\mu(\hat\vartheta)\big)\) as an estimator of \(\boldsymbol\Sigma(\vartheta)\), based on methods from the generalized estimating equation (GEE) literature \autocite{kauermann2001note}.
\textcite{ozenne2020small} show that asymptotic bias of $\hat{\boldsymbol\Sigma}$ is a negative definite matrix, implying that $\hat{\boldsymbol\Sigma}$ tends to underestimate $\boldsymbol\Sigma(\vartheta)$.
Based on this observation they devise an iterative scheme that use the ML estimator of \(\boldsymbol\Lambda\) and \(\mathbf B\), and aims to correct the bias in the estimation of the variance parameters \(\boldsymbol\Psi\) and \(\boldsymbol\Theta\), by correcting the bias of the \(\hat{\boldsymbol\Sigma}\) as an estimator of \(\boldsymbol\Sigma(\vartheta)\).
Importantly, that method focuses only on the variance parameters in the model, and the bias of the estimators of regression coefficients and factor loadings are not targeted.

\subsection{Reduced-Bias M-Estimation}\label{sec-rbm}

Reduced-bias M-estimation \autocite[RBM-estimation,][]{kosmidis2024empirical} is a recently-introduced, general-purpose bias reduction method that can operate implicitly or explicitly. 
Approximating the bias function requires up to the third derivatives of the log-likelihood, but in practice only the log-likelihood function, its gradient, and its Hessian are needed.
As a result, RBM estimation can be readily implemented, either using closed-form derivatives, or through numerical differentiation routines, such as those provided by the \texttt{\{numDeriv\}} R package \autocite{gilbert2022numderiv}.

When estimation involves maximizing a log-likelihood, as is the case for maximum likelihood estimation of SEMs, the implicit RBM (iRBM) estimator $\tilde\vartheta$ is defined as the maximizer of the penalized log-likelihood $\ell(\vartheta) + P(\vartheta)$, where
\begin{equation}
P(\vartheta) =
-\frac{1}{2}\operatorname{tr}\big\{j(\vartheta)^{-1} e(\vartheta)\big\} \label{eq-penobjfun}
\end{equation}
with \(j(\vartheta)\) the negative Hessian matrix \(-\nabla \nabla^\top \ell(\vartheta)  \in \mathbb R^{m\times m}\) and \(e(\vartheta)\) an $m\times m$ (symmetric) matrix whose \((k, l)\)th element is
\[ 
[e(\vartheta)]_{kl} = \sum_{i=1}^n \left\{
\frac{\partial \ell_i(\vartheta)}{\partial \vartheta_l} \frac{\partial
\ell_i(\vartheta)}{\partial \vartheta_k} \right\} \, .
\]
Starting with the ML estimator $\hat\vartheta$, the explicit RBM (eRBM) estimator is defined as
\[
\vartheta^\dagger = \hat{\vartheta} + j(\hat{\vartheta})^{-1} A(\hat{\vartheta}) \,,
\]
where \(A(\hat\vartheta) = \nabla P(\vartheta)\big|_{\vartheta=\hat\vartheta} \in \mathbb R^m\).
The RBM estimators have the same asymptotic distribution as the original estimators, ensuring that standard inference and model selection procedures apply directly.

To appreciate the role of $P(\vartheta)$, it is helpful to view RBM estimation as beginning with the goal of solving a bias-corrected score equation and then asking: what modification to the log-likelihood would yield this solution? 
In this sense, RBM works backward from the adjusted score equations to define a penalized likelihood whose maximizer removes the leading $O(n^{-1})$ bias term from the estimator. 
The resulting penalty $P(\vartheta)$ has the appealing property that its gradient $A(\vartheta)=\nabla P(\vartheta)$ provides a sample-based adjustment that, on average, cancels the first-order bias. 
Intuitively, $P(\vartheta)$ ``tilts'' the log-likelihood just enough---based on the observed data---to shift the score equation’s root toward the bias-reduced solution, without inflating the estimator’s $O(n^{-1/2})$ variance.
For full details, the reader is invited to consult Section~5 of \textcite{kosmidis2024empirical}.

The practical implementation of RBM within the normal SEM framework requires translating the general quantities in \eqref{eq-penobjfun} into expressions under the multivariate normal likelihood with model-implied mean vector $\boldsymbol\mu(\vartheta)$ and covariance matrix $\boldsymbol\Sigma(\vartheta)$ given by \eqref{eq-marg-mean} and \eqref{eq-marg-var}.
Thus, the penalty components depend intricately on the derivatives of $\boldsymbol\mu(\vartheta)$ and $\boldsymbol\Sigma(\vartheta)$ with respect to the free parameters $\vartheta$ of the SEM.
We derive analytic closed-form expressions of these first-order derivatives via the chain rule for a general SEM in Appendix \ref{sec-derivatives}.
Higher-order derivatives required to evaluate $\nabla P(\vartheta)$ are obtained by numerical differentiation of these first-order quantities, as is common in SEM software such as \texttt{\{lavaan\}} \citep{rosseel2012lavaan}.
This strategy facilitates a robust application to a wide array of SEM specifications.

Explicit and implicit RBM estimation have advantages and disadvantages.
Computing \(\vartheta^\dagger\) is a single step procedure, so is likely to be faster to obtain than \(\tilde{\vartheta}\).
However, the explicit approach provides no safeguards on what value the final estimator can take, and it may result in an estimate outside the parameter space.
The implicit approach allows appropriate constraints to be incorporated into the optimisation process, and is typically more stable as it does not require the value of \(\hat{\vartheta}\).
Overall, both explicit and implicit RBM estimation should be noticeably faster than simulation-based methods such as the bootstrap and jackknife, because it does not require repeated fits.

\section{Simulation Study}\label{sec-simulation}

We evaluate the effectiveness of explicit and implicit RBM estimation through an extensive simulation study that uses the same models and experimental settings as \textcite{dhaene2022resampling}.
Their study compared SEM adaptations of the resampling-based methods for bias reduction in Section~\ref{sec-resampling} to the alternative methods for bias correction in Section~\ref{sec-alternative}.
By adding eRBM and iRBM to this simulation design, we can perform a side-by-side comparison of the performance of RBM estimation with previously proposed methods.

\subsection{Models}\label{models}

Two distinct SEMs are considered: 
(a) A two-factor SEM with a latent regression; and 
(b) a latent growth curve model (GCM).
Both models represent commonly-used models by practitioners and researchers in the social sciences and beyond.
Below we give the details of these two models, in particular their parameterization and the conditions under which they are simulated.

As the name implies, the two-factor SEM contains two latent variables, \(\eta_{1}\) and \(\eta_{2}\).
Each latent variable is measured by three observed variables: 
\(y_{1}\), \(y_{2}\), \(y_{3}\) for \(\eta_{1}\) and \(y_{4}\), \(y_{5}\), \(y_{6}\) for \(\eta_{2}\).
The structural model is given by a single regression of $\eta_{1}$ on $\eta_{2}$.
Thus, the loading matrix and structural coefficient matrix take the form
$$
\boldsymbol \Lambda = \begin{bmatrix}
\lambda_{11} & 0 \\
\lambda_{21} & 0 \\
\lambda_{31} & 0 \\
0 & \lambda_{42} \\
0 & \lambda_{52} \\
0 & \lambda_{62} \\
\end{bmatrix} \quad \text{and} \quad
\mathbf B = \begin{bmatrix}
0 & 0 \\
\beta & 0 \\
\end{bmatrix} \, ,
$$
while \(\boldsymbol\Theta\) and \(\boldsymbol\Psi\) are diagonal matrices with \(\theta_{11}, \ldots, \theta_{66}\) and \(\psi_{11}, \psi_{22}\) on their diagonals, respectively.
The item intercepts $\boldsymbol\nu=(\nu_1,\dots,\nu_6)$ are freely estimable, but the factor intercepts $\boldsymbol\alpha$ are fixed to zero as the location of the latent variables cannot be identified otherwise.
Scale identification is achieved by fixing $\lambda_{11} = \lambda_{42} = 1$, thereby anchoring the measurement scale of the latent variables.

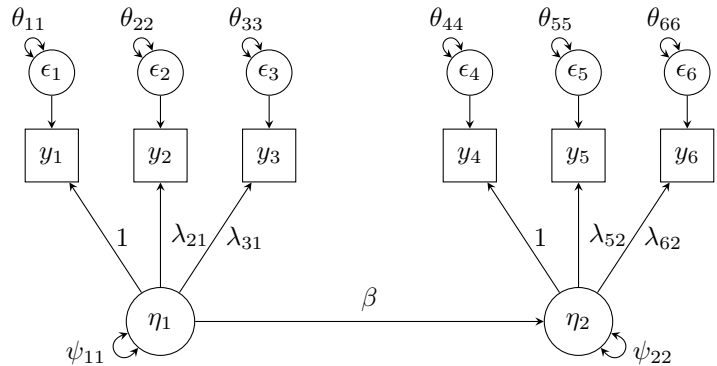
\begin{figure}[htbp]
\centering
\begin{tikzpicture}[%
  scale=1.1,
  >=stealth,
  auto,
  node distance=1.6cm,
  every node/.style={font=\small},
  latent/.style={circle, draw, minimum size=9mm, inner sep=0mm},
  obs/.style={rectangle, draw, minimum size=7mm, inner sep=0mm},
  error/.style={circle, draw, minimum size=6mm, inner sep=0mm}
]

\node[latent] (eta1) at (-1.5, -2) {$\eta_{1}$};
\node[latent] (eta2) at ( 3.5, -2) {$\eta_{2}$};

\node[obs] (y1) at (-2.8, 0) {$y_{1}$};
\node[obs] (y2) at (-1.5, 0) {$y_{2}$};
\node[obs] (y3) at ( -0.2, 0) {$y_{3}$};

\node[obs] (y4) at ( 2.2, 0) {$y_{4}$};
\node[obs] (y5) at ( 3.5, 0) {$y_{5}$};
\node[obs] (y6) at ( 4.8, 0) {$y_{6}$};

\node[error] (e1) at (-2.8, 1) {$\epsilon_{1}$};
\node[error] (e2) at (-1.5, 1) {$\epsilon_{2}$};
\node[error] (e3) at ( -0.2, 1) {$\epsilon_{3}$};
\node[error] (e4) at ( 2.2, 1) {$\epsilon_{4}$};
\node[error] (e5) at ( 3.5, 1) {$\epsilon_{5}$};
\node[error] (e6) at ( 4.8, 1) {$\epsilon_{6}$};

\draw[->] (e1) -- (y1);
\draw[->] (e2) -- (y2);
\draw[->] (e3) -- (y3);
\draw[->] (e4) -- (y4);
\draw[->] (e5) -- (y5);
\draw[->] (e6) -- (y6);

\draw[->] (eta1) -- node[pos=0.5, right] {$1$} (y1);
\draw[->] (eta1) -- node[pos=0.5, right] {$\lambda_{21}$} (y2);
\draw[->] (eta1) -- node[pos=0.5, right] {$\lambda_{31}$} (y3);

\draw[->] (eta2) -- node[pos=0.5, right] {$1$} (y4);
\draw[->] (eta2) -- node[pos=0.5, right] {$\lambda_{52}$} (y5);
\draw[->] (eta2) -- node[pos=0.5, right] {$\lambda_{62}$} (y6);

\draw[->] (eta1) -- node[midway, above] {$\beta$} (eta2);

\draw[<->] (eta1) to[out=200, in=230, looseness=4] 
  node[left] {$\psi_{11}$} (eta1);

\draw[<->] (eta2) to[out=-50, in=-20, looseness=4] 
  node[right] {$\psi_{22}$} (eta2);

\foreach \i in {1,...,6}{
  \draw[<->] (e\i)
    to[out=110, in=140, looseness=5]
    node[above] {$\theta_{\i\i}$}
    (e\i);
}

\end{tikzpicture}
\caption{Path diagram of the two-factor SEM used in the simulation study, shown at the population level (suppressing individual subscripts). Each factor is measured by three indicators, with factor intercepts are fixed to zero and scale identification set by fixing $\lambda_{11}=\lambda_{42}=1$. The structural part specifies a single regression of $\eta_{1}$ on $\eta_{2}$.}
\label{fig-path-twofac}
\end{figure}

The corresponding path diagram is shown in Figure~\ref{fig-path-twofac}.
In total there are 19 estimable parameters.
Although, from \eqref{eq-sem_lik} and also the score equation \eqref{eq-score-mu}, it is clear that the ML estimator for $\boldsymbol\nu$ is simply the sample mean vector $\bar{\mathbf y}$.
Given this, only 13 parameters remain to be estimated after centring the data.

\phantomsection\label{cell-fig-path-twofac}

The latent GCM captures longitudinal change in continuous outcomes $y_{1},\dots,y_{10}$ using two latent factors: 
$\eta_1$ for the \emph{intercept} (average initial status) and $\eta_2$ for the \emph{slope} (linear growth rate).
This time, the latent intercepts $\boldsymbol\alpha = (\alpha_1, \alpha_2)^\top$ are freely estimated to provide the population mean initial status and average rate of change, respectively.
Each observed variable loads on both factors in a cross-loading design.
All loadings on the intercept factor are fixed to one to anchor the factor’s scale and interpret it as the common level across all measurement occasions.
Loadings on the slope factor are fixed to $0,1,\dots,9$ to represent equally spaced measurement occasions starting at 0, making $\alpha_1$ directly interpretable as the expected outcome at the first measurement occasion.
Therefore, in this model, $\boldsymbol\Lambda$ and $\mathbf B$ are fixed quantities:
$$
\boldsymbol\Lambda = \begin{bmatrix}
1 & 0 \\
1 & 1 \\
\vdots & \vdots \\
1 & 9 \\
\end{bmatrix} \quad \text{and} \quad 
\mathbf B = \begin{bmatrix}
0 & 0 \\
0 & 0 \\
\end{bmatrix}.
$$ 

The residual covariance matrix is specified as a scalar multiple of the identity $\boldsymbol\Theta = \theta_{11} \mathbf I$ so that all error variances are equal across measurement occasions.
This choice reflects the common assumption---under the multilevel interpretation of the latent GCM as a random intercept/random slope model---that repeated measures are equally reliable at each time point.
Now, $\mathbf y_i \mid \boldsymbol \eta_i \sim \text{N}_p(\boldsymbol\Lambda \boldsymbol\alpha + \boldsymbol\Lambda
\boldsymbol\eta, \theta_{11} \mathbf I)$ with \(\boldsymbol\eta_i \sim \text{N}_2(\mathbf 0, \boldsymbol\Psi)\), and stacking the per-person response vectors yields the equivalent linear mixed effects representation, and REML (Section~\ref{sec-alternative}) can be used for the estimation of the variance parameters.
All that remains is to specify the latent covariance matrix $\boldsymbol\Psi$, a $2\times 2$ symmetric matrix with diagonal elements $\psi_{11}$ and $\psi_{22}$ and off-diagonal element $\psi_{12}$, all of which are freely estimated.

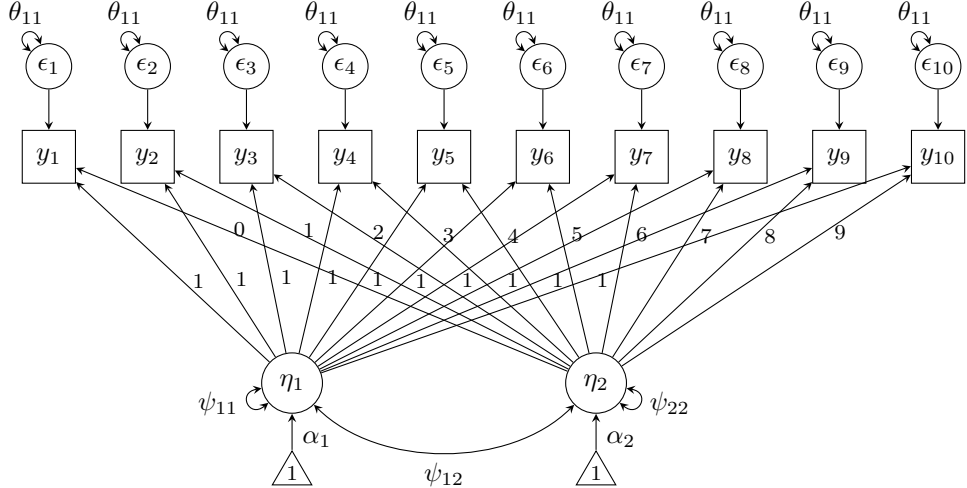
\begin{figure}[htbp]
\centering
\usetikzlibrary{shapes.geometric,arrows,calc,positioning}
\begin{tikzpicture}[%
  >=stealth,              
  auto,                   
  node distance=1.6cm,
  every node/.style={font=\small},
  latent/.style={circle, draw, minimum size=8mm, inner sep=0mm},
  obs/.style={rectangle, draw, minimum size=7mm, inner sep=0mm},
  error/.style={circle, draw, minimum size=6mm, inner sep=0mm},
intercept/.style={regular polygon, regular polygon sides=3, draw, inner sep=0pt, minimum size=6mm}
  ]

\node[latent] (i) at (-2,-3) {$\eta_{1}$};
\node[latent] (s) at ( 2,-3) {$\eta_{2}$};
\node[intercept] (int1) at (-2,-4.2) {\footnotesize 1};
\node[intercept] (int2) at (2,-4.2) {\footnotesize 1};

\draw[<->] (i) to[out=-45, in=225,looseness=0.9] node[below] {$\psi_{12}$} (s);

\draw[<->] (i) to[out=190, in=220, looseness=4] 
  node[left] {$\psi_{11}$} (i);

\draw[<->] (s) to[out=-40, in=-10, looseness=4] 
  node[right] {$\psi_{22}$} (s);

\draw[->] (int1) -- node[right,pos=0.3] {$\alpha_1$} (i);
\draw[->] (int2) -- node[right,pos=0.3] {$\alpha_2$} (s);

\foreach \j in {1,...,10} {
  \pgfmathsetmacro{\x}{(\j*1.3 - 6.5)}

  \node[obs] (y\j) at (\x,0) {$y_{\j}$};

  \node[error] (e\j) at (\x,1.2) {$\epsilon_{\j}$};

  \draw[->] (e\j) -- (y\j);

  \draw[->] (i) -- node[pos=0.45,right] {\footnotesize 1} (y\j);

  \pgfmathtruncatemacro{\loading}{\j - 1}
  \draw[->] (s) -- node[pos=0.7,right] {\footnotesize\loading} (y\j);
}

\foreach \i in {1,...,10}{
  \draw[<->] (e\i)
    to[out=110, in=140, looseness=5]
    node[above] {$\theta_{11}$}
    (e\i);
}

\end{tikzpicture}
\caption{Path diagram of the latent GCM used in the simulation study, shown here at the population level (suppressing individual subscripts). Each observed outcome loads on intercept and slope factors, with intercept loadings fixed to 1 and slope loadings fixed to $0,1,\dots,9$ to represent equally spaced measurement occasions. Residual variances are constrained equal across time, and $\boldsymbol\alpha$ and $\boldsymbol\Psi$ are freely estimated.}
\label{fig-path-growth}
\end{figure}

The total number of parameters in this latent GCM, with all the constraints, is six.
This is identical to the number of parameters estimable in the linear mixed effects representation.
The corresponding path diagram is shown in Figure~\ref{fig-path-growth}.

\phantomsection\label{cell-fig-path-growth}

\subsection{Simulation Settings}\label{simulation-settings}

We assess the robustness and performance of RBM estimation, by considering all possible combinations of three experimental conditions.
The first condition is the sample size, taking values $n \in \{15, 20, 50, 100, 1000\}$.
The focus is on small sample sizes $n\leq 100$, but $n=1000$ serves as a large-sample benchmark for assessing limiting bias behaviour.

The second condition is the average reliability of the items, which is defined as 
\[
\operatorname{Rel}(\vartheta) = \frac{1}{p}\sum_{j = 1}^p \frac{\boldsymbol\Sigma^*_{jj}(\vartheta)}{\boldsymbol\Sigma_{jj}(\vartheta)} \, .
\] 
with \(\boldsymbol\Sigma^*(\vartheta)\) and \(\boldsymbol\Sigma(\vartheta)\) as in (\ref{eq-marg-var}).
We consider two sets of parameters as per Table~\ref{tbl-truth-twofac} for the two-factor SEM, and the two sets of parameters as per Table~\ref{tbl-truth-growth} for the latent GCM, which result in reliability being roughly either 0.5 (low) or 0.8 (high).
Low reliability means that much of the variance in the data is attributable to measurement error, which makes estimation of the variance components more challenging and can degrade both estimation accuracy and inferential performance.

\begin{table}[htbp]
\centering
\setlength{\tabcolsep}{6pt}

\begin{subtable}[t]{0.48\linewidth}
\centering
\caption{Two-factor SEM parameters}
\label{tbl-truth-twofac}
\begin{tabular}{lcc}
\toprule
$\vartheta$ & $\operatorname{Rel}(\vartheta)=0.8$ & $\operatorname{Rel}(\vartheta)=0.5$ \\
\midrule\addlinespace[2.5pt]
$\lambda_{21}$ & 0.7 & 0.7 \\
$\lambda_{31}$ & 0.6 & 0.6 \\
$\lambda_{52}$ & 0.7 & 0.7 \\
$\lambda_{62}$ & 0.6 & 0.6 \\
$\beta$        & 0.25 & 0.25 \\
$\theta_{11}$ & 0.25 & 1 \\
$\theta_{22}$ & 0.1225 & 0.49 \\
$\theta_{33}$ & 0.09 & 0.36 \\
$\theta_{44}$ & 0.25 & 1 \\
$\theta_{55}$ & 0.1225 & 0.49 \\
$\theta_{66}$ & 0.09 & 0.36 \\
$\psi_{11}$    & 1 & 1 \\
$\psi_{22}$    & 1 & 1 \\
\bottomrule
\end{tabular}
\end{subtable}
\hfill
\begin{subtable}[t]{0.48\linewidth}
\centering
\caption{Latent GCM parameters}
\label{tbl-truth-growth}
\begin{tabular}{lcc}
\toprule
$\vartheta$ & $\operatorname{Rel}(\vartheta)=0.8$ & $\operatorname{Rel}(\vartheta)=0.5$ \\
\midrule\addlinespace[2.5pt]
$\alpha_1$  & 0   & 0    \\
$\alpha_2$  & 0   & 0    \\
$\psi_{11}$ & 550 & 275  \\
$\psi_{22}$ & 100 & 50   \\
$\psi_{12}$ & 40  & 20   \\
$\theta_{11}$  & 500 & 1300 \\
\bottomrule
\end{tabular}
\end{subtable}

\caption{Simulation parameters under two reliability settings.}
\label{tbl:side_by_side}
\end{table}

\begin{figure}[htbp]

\centering{

\includegraphics[width=1\linewidth,height=\textheight,keepaspectratio]{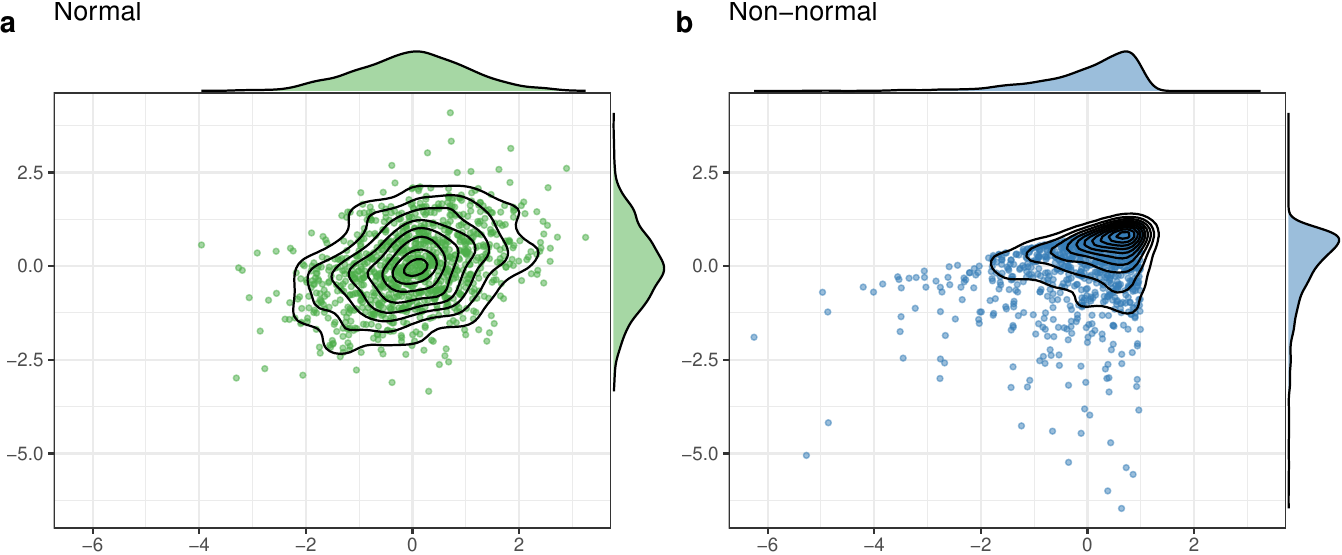}

}

\caption{\label{fig-compare-dist}Scatterplots illustrating the distributional conditions used in the simulation study. Each plot displays 1000 randomly generated observations drawn from a distribution with zero mean vector and covariance matrix \(\begin{bmatrix} 1 & 0.3 \\ 0.3 & 1 \end{bmatrix}\). Marginal density plots on the axes highlight differences in symmetry and tail behaviour.}

\end{figure}

The third condition is the distributional assumption of the model, which we set to be either ``normal'' or ``non-normal'', for both the latent factors $\eta$ and the measurement errors $\epsilon$.
We simulate datasets to have the covariance matrix (\ref{eq-marg-var}), as dictated by the parameters settings in Table~\ref{tbl-truth-twofac} and Table~\ref{tbl-truth-growth}, for the two-factor SEM and the latent GCM, respectively, but determine the distribution of the latent factors and the measurement errors using the \texttt{\{covsim\}} R package \autocite{gronneberg2022covsim}.
This package generates multivariate data with a target covariance matrix while controlling excess kurtosis and skewness of the distribution.
In the case of the normal distribution, both the skewness and excess kurtosis are set to \(0\).
In the non-normal case, skewness is set to \(-2\) with excess kurtosis of \(6\), producing asymmetric, heavy-tailed distributions.
Figure~\ref{fig-compare-dist} illustrates the shapes of the distributions for normal and non-normal distributions.

\phantomsection\label{cell-fig-compare-dist}

\subsection{Performance Metrics}\label{sec-perf}

Across the \(5 \times 2 \times 2 = 20\) distinct experimental settings for sample size \(n\), reliability \(\operatorname{Rel}(\vartheta)\), and distributional assumptions, we simulate 2000 independent datasets for each of the two-factor SEM and latent GCM.
Each simulated dataset is then analyzed using ML, eRBM, and iRBM estimation, as well as the bias-correction method of \textcite[Sec. 4]{ozenne2020small}, and the bootstrap and jackknife adaptations proposed by \textcite{dhaene2022resampling}.
Due to the equivalence of the latent GCM model with a linear mixed effects model, REML is also used to estimate the latent GCM model (but not the two-factor model).

The performance of each method is first evaluated in terms of the number of samples that result in acceptable estimates, denoted $R$.
We define acceptable estimates as those that result 
(a) from successful convergence of the optimisation procedure that computes them; 
(b) in positive definite \(\boldsymbol\Sigma(\vartheta)\) at the estimates; and 
(c) in the diagonal elements of \(j(\vartheta)^{-1}\) at the estimates being within acceptable numerical ranges.
The last of these criteria serves as a diagnostic to exclude estimates with extreme or unreliable parameter variances.
For the two-factor SEM and the latent GCM, we require all standard errors to be less than 5 and 500, respectively, reflecting differences in the scaling of observed variables.

Let $\dot{\vartheta}^{(r)}_j$ be the \(j\)th component of the vector of estimates from the \(r\)th iteration, $r=1,\dots,R$, that resulted in acceptable estimates.
We evaluate the methods in terms of simulation-based estimates of
(a) mean bias $R^{-1} \sum_{r=1}^{R}(\dot{\vartheta_j}^{(r)} - \vartheta_j)$;
(b) probability of underestimation (PU) $\sum_{r=1}^{R} \mathbbm 1 [\dot{\vartheta}^{(r)}_j < \vartheta_j]/R$; and 
(c) root mean squared error (RMSE) \(\sqrt{\sum_{r=1}^{R}(\dot{\vartheta}^{(r)}_j - \vartheta_j)^2 / R}\).
We aim to demonstrate that the proposed eRBM and iRBM estimators outperform ML by exhibiting smaller bias and RMSE, as well as probabilities of underestimation closer to 0.5, indicating reduced median bias.
To provide a like-for-like comparison across all methods, we also reproduce the \emph{relative} mean bias and RMSE plots (Figures 4, 5, 7 \& 8) of \textcite[][]{dhaene2022resampling}, augmenting them with our eRBM and iRBM results for a clear evaluation.
The formula to calculate relative mean bias is $\sum_{r=1}^{R}\dot{\vartheta_j}^{(r)}/(R\vartheta_j) - 1$.

Despite the fact that bias-reduction techniques are not primarily intended to improve inferential performance, we also evaluate each method using the simulation-based estimate of the conditional coverage of \(95\%\) Wald-type confidence intervals at the estimates 
$$
\frac{1}{R}\sum_{r=1}^{R} \mathbbm 1  \left[\dot{\vartheta}^{(r)}_j - 1.96 s^{(r)}_j \leq \vartheta_j \leq \dot{\vartheta}^{(r)}_j + 1.96 s^{(r)}_j\right] \, ,
$$
where \(s_j^{(r)}\) is the estimated standard error of \(\dot\vartheta^{(r)}_r\).
As mentioned in Section~\ref{sec-rbm}, RBM estimators have the same asymptotic distribution as the ML estimator.
Hence, we can obtain robust standard errors by using the variance-covariance matrix of the asymptotic distribution of the ML estimator under model misspecification, as given in \eqref{eq-robustasymp}.
Specifically, \(s^{(r)}_j\) is taken to be the square root of the \(j\)th diagonal element of \(j(\dot\vartheta^{(r)})^{-1}
e(\dot\vartheta^{(r)}) j(\dot\vartheta^{(r)})^{-1}\) for ML, eRBM, iRBM, and the method in \textcite[Sec. 4]{ozenne2020small}. 
For bootstrap and jackknife, standard errors \(s^{(r)}_j\) are computed as the standard deviation of the replicate estimates, and 95\% confidence intervals are constructed as $\dot\vartheta_j \pm 1.96 s^{(r)}_j$ using the normal approximation \autocite[Sec. 8.3][]{wasserman2004all}.
We do not consider alternative internals for bootstrap, like basic, studentized, percentile and BCa intervals \autocite[see,][for details]{davison1997bootstrap} as they have been found to perform similarly to the normal approximation interval in the experimental settings we consider here \autocite[see][]{dhaene2022resampling}.
For REML, standard errors are computed from the inverse of the observed Hessian of the REML log-likelihood.

\subsection{Computational Note}

All analyses were conducted in R \autocite{R2025}.
Computations were performed on a 15-inch MacBook Air (2023) equipped with an Apple M2 chip (8-core CPU, 4 performance cores up to 3.50 GHz) and 16 GB of unified memory.

ML estimation, including robust standard errors, was performed using the \texttt{\{lavaan\}} package \autocite[with \texttt{se = "robust.huber.white"},][]{rosseel2012lavaan}. 
The eRBM and iRBM estimators and robust standard errors were implemented using our freely available \texttt{\{brlavaan\}} package, which also provides R functions for the \textcite{ozenne2020small} correction, bootstrap, and jackknife procedures tailored to the two-factor SEM and latent GCM considered here. 
The R package \texttt{\{lme4\}} \autocite{bates2015fitting} was used for fitting the latent GCM as a LMM.
All code used to perform the simulations, along with the computed simulation results, is openly available on the Open Science Framework repository via the URL provided in the supplementary materials.

Finally, for comparability with the prior study of \textcite{dhaene2022resampling}, we employ bounded estimation (in \texttt{\{lavaan\}} and \texttt{\{brlavaan\}}, use option \texttt{bounded = "standard"}) as described in \textcite{dejonckere2022using}. 
This approach applies data-informed upper and lower parameter bounds within the constrained optimisation routine and was particularly helpful for the small-sample settings ($n < 100$) considered here, where it greatly reduced convergence failures and the need to discard replications. 

\section{Results}\label{results}

\subsection{Rates of acceptable estimates}\label{rates-of-acceptable-estimates}

For ML estimation, all parameter estimates were acceptable across every experimental setting for both models, including REML estimation of the latent GCM.
Table~\ref{tbl-convsucc} reports the rates of acceptable estimates for eRBM and iRBM. 
For the two-factor model, the most problematic scenarios were those with $n \leq 20$ in the low-reliability setting, regardless of outcome distribution. 
Here, eRBM produced markedly lower rates of acceptable estimates, with a minimum of $39.3\%$ reported, whereas iRBM performed considerably better, achieving rates of $89.5\%$ and above even in these challenging scenarios. 
For the latent GCM, no drastic issues were observed with eRBM or iRBM, with acceptance rates mostly at or above $82\%$, which is substantially higher than the minimum rates observed for eRBM in the two-factor SEM.

Since the ML estimates have been acceptable in all samples, the unacceptable eRBM estimates are due to the post hoc correction resulting in a non-positive definite estimate of \(\boldsymbol\Sigma(\vartheta)\) or excessively large estimated standard errors.
As for iRBM, the vast majority of failures were due to unsuccessful convergence of the \texttt{nlminb()} R method that was used for optimisation. 
Trying different starting values, or a set of candidate optimisation procedures on each sample, and keeping the estimates from those that succeed, could substantially increase the rate of acceptable estimates for iRBM.

\begin{table}[htbp]

\caption{\label{tbl-convsucc}Rate of acceptable estimates (percentage) for estimation of both models using eRBM, iRBM and bootstrap. Scenarios with success rates below 50\% are highlighted in shades of red, with varying intensity signifying severity.}

\centering
\fontsize{9pt}{10.8pt}\selectfont
\setlength{\tabcolsep}{4pt}  

\begin{tabular*}{\linewidth}{@{\extracolsep{\fill}}rrrrrrrrrrrrr}
\toprule
 & \multicolumn{6}{c}{Two-factor SEM} & \multicolumn{6}{c}{Growth curve model} \\ 
\cmidrule(lr){2-7} \cmidrule(lr){8-13}
 & \multicolumn{3}{c}{Reliability = 0.8} & \multicolumn{3}{c}{Reliability = 0.5} & \multicolumn{3}{c}{Reliability = 0.8} & \multicolumn{3}{c}{Reliability = 0.5} \\ 
\cmidrule(lr){2-4} \cmidrule(lr){5-7} \cmidrule(lr){8-10} \cmidrule(lr){11-13}
$n$ & eRBM & iRBM & boot & eRBM & iRBM & boot & eRBM & iRBM & boot & eRBM & iRBM & boot \\ 
\midrule\addlinespace[2.5pt]
\multicolumn{13}{l}{Normal} \\[2.5pt] 
\midrule\addlinespace[2.5pt]
15 & 91.2 & 98.4 & \cellcolor[HTML]{FDB4A1}{{32.6}} & \cellcolor[HTML]{FFE0D7}{{43.0}} & 96.9 & \cellcolor[HTML]{FDB3A0}{{32.3}} & 100 & 95.6 & \cellcolor[HTML]{CD0000}{\textcolor[HTML]{FFFFFF}{0.0}} & 100 & 98.7 & \cellcolor[HTML]{CD0000}{\textcolor[HTML]{FFFFFF}{0.0}} \\ 
20 & 96.7 & 99.2 & 100.0 & 51.9 & 98.1 & 100.0 & 100 & 98.2 & \cellcolor[HTML]{CD0000}{\textcolor[HTML]{FFFFFF}{0.0}} & 100 & 99.6 & \cellcolor[HTML]{CD0000}{\textcolor[HTML]{FFFFFF}{0.0}} \\ 
50 & 100.0 & 100.0 & 100.0 & 91.4 & 99.8 & 100.0 & 100 & 99.9 & 98.0 & 100 & 100.0 & 96.1 \\ 
100 & 100.0 & 100.0 & 100.0 & 99.9 & 100.0 & 100.0 & 100 & 100.0 & 98.0 & 100 & 100.0 & 100.0 \\ 
1000 & 100.0 & 100.0 & 100.0 & 100.0 & 100.0 & 100.0 & 100 & 100.0 & 100.0 & 100 & 100.0 & 100.0 \\ 
\midrule\addlinespace[2.5pt]
\multicolumn{13}{l}{Non-normal} \\[2.5pt] 
\midrule\addlinespace[2.5pt]
15 & 81.2 & 89.5 & \cellcolor[HTML]{FBAC98}{{30.7}} & \cellcolor[HTML]{FFD0C4}{{39.3}} & 86.0 & \cellcolor[HTML]{FCB19E}{{31.9}} & 100 & 81.9 & \cellcolor[HTML]{CD0000}{\textcolor[HTML]{FFFFFF}{0.0}} & 100 & 95.2 & \cellcolor[HTML]{CD0000}{\textcolor[HTML]{FFFFFF}{0.0}} \\ 
20 & 90.6 & 94.8 & 99.0 & \cellcolor[HTML]{FFFEFD}{{50.2}} & 91.0 & 100.0 & 100 & 86.9 & \cellcolor[HTML]{CD0000}{\textcolor[HTML]{FFFFFF}{0.0}} & 100 & 97.4 & \cellcolor[HTML]{CD0000}{\textcolor[HTML]{FFFFFF}{0.0}} \\ 
50 & 99.9 & 99.8 & 100.0 & 85.7 & 98.0 & 100.0 & 100 & 98.2 & 94.2 & 100 & 99.8 & 77.9 \\ 
100 & 100.0 & 100.0 & 100.0 & 99.1 & 100.0 & 100.0 & 100 & 100.0 & 96.1 & 100 & 100.0 & 93.2 \\ 
1000 & 100.0 & 100.0 & 100.0 & 100.0 & 100.0 & 100.0 & 100 & 100.0 & 100.0 & 100 & 100.0 & 100.0 \\ 
\bottomrule
\end{tabular*}

\end{table}%

For completeness, we report that nearly all jackknife and \textcite{ozenne2020small} corrections yielded acceptable estimates, with minimum rates of $98.9\%$ and $99.3\%$, respectively, across all settings we considered; consistent with the reporting in \textcite[Sec.~6]{dhaene2022resampling}. 
The acceptable rates for the bootstrap on the other hand are worth dissecting, and for context, have been added to Table~\ref{tbl-convsucc}.
Note that the bootstrap rates are based on the summaries in Table 1 of \textcite{dhaene2022resampling}.
Bootstrap performance was particularly poor in the $n = 15$ and $n = 20$ settings, achieving $32.3\%$ or less across all conditions in this sample size range---substantially worse than either eRBM or iRBM.
As a post hoc correction, bootstrap can also result in non-positive definite estimates of \(\boldsymbol\Sigma(\vartheta)\) or excessively large estimated standard errors, but, in addition, its rates are impacted by unacceptable ML estimates from bootstrap samples.

It is important to clarify that a reported bootstrap acceptance rate of $0\%$ does not imply that all bootstrap results were discarded. 
Our definition of ``unacceptable'' is stricter than that of \textcite{dhaene2022resampling}: 
we classified a bootstrap replication as unacceptable if \emph{any} of its  resamples were problematic, whereas \textcite{dhaene2022resampling} simply excluded the unsuccessful resamples and retained the remaining ones. 
They further trimmed point estimates exceeding the 0.005 and 0.995 quantiles to mitigate the effect of outliers. 
In the end, our analysis used the same bootstrap results as \textcite{dhaene2022resampling}, but we report these stricter acceptance rates to highlight the large number of failed resamples that will likely occur in practice.

\subsection{Two-Factor SEM: Bias, RMSE, and Coverage}\label{two-factor-sem}

\begin{figure}[htbp]
\centering
\includegraphics[width=1\linewidth,height=\textheight,keepaspectratio]{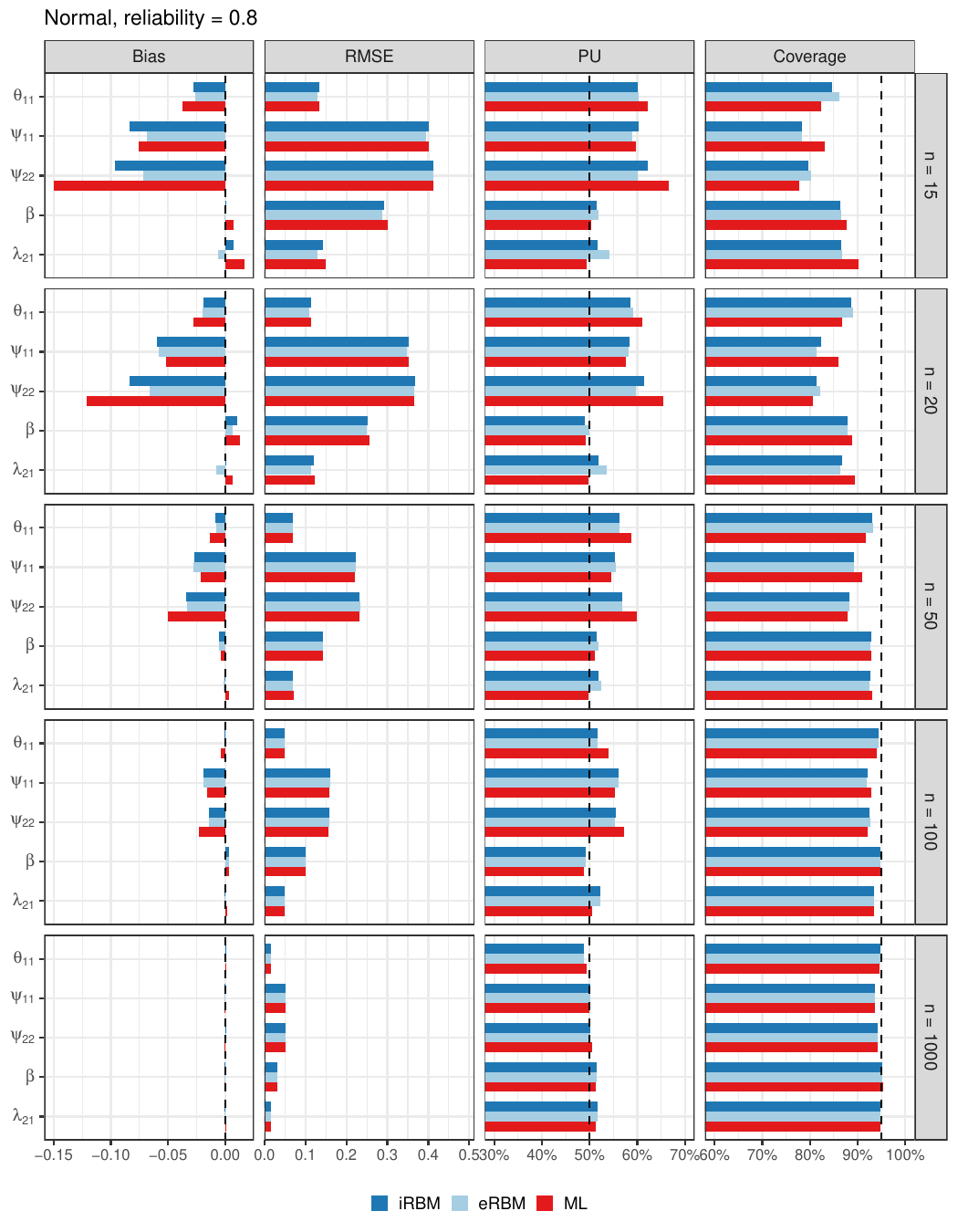}
\caption{\label{fig-perf-twofac}Performance metrics (bias, RMSE, PU, and coverage) of the ML, eRBM, and iRBM estimators for the two-factor SEM. Vertical dashed lines indicate the ideal values for each metric.}
\end{figure}%

\begin{figure}[p]
\centering

\begin{subfigure}{\linewidth}
  \centering
  \includegraphics[width=\linewidth]{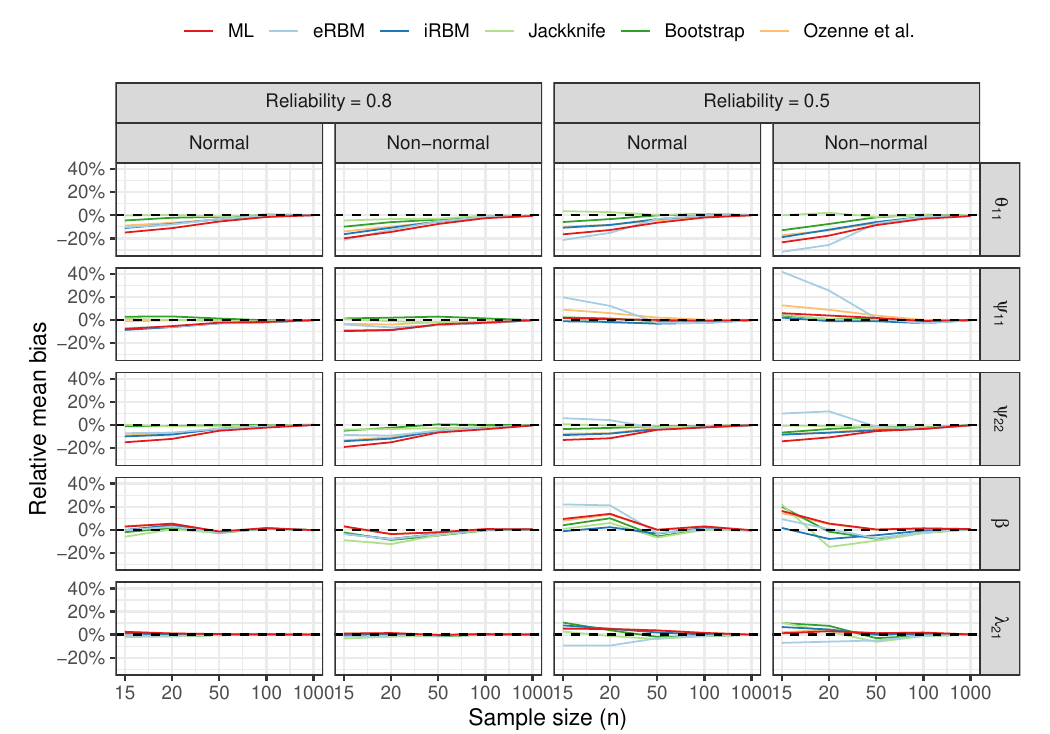}
  \caption{Relative mean bias.}
  \label{fig:twofac-bias}
\end{subfigure}

\vspace{1em} 

\begin{subfigure}{\linewidth}
  \centering
  \includegraphics[width=\linewidth]{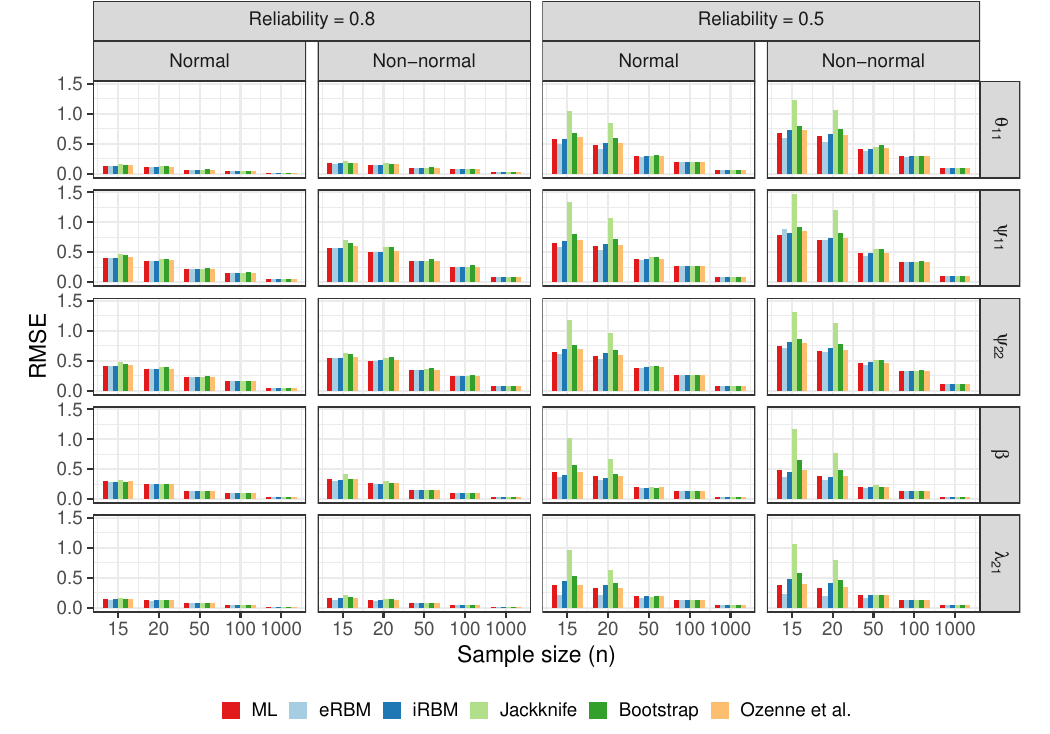}
  \caption{Root mean squared error (RMSE).}
  \label{fig:twofac-rmse}
\end{subfigure}

\caption{\label{fig-twofac}
Comparison of the ML, eRBM, and iRBM estimators with jackknife, bootstrap, and Ozenne corrections for estimation of the two-factor SEM.}
\end{figure}

Similar to \textcite{dhaene2022resampling}, our analysis of the two-factor model focuses on a subset of key parameters: 
the error variance for item $y_1$ ($\theta_{11}$), 
the second factor loading for item $y_1$ ($\lambda_{21}$), 
the two latent factor variances ($\psi_{11}$ and $\psi_{22}$), 
and the regression coefficient $\beta$. 
This selection spans a range of parameter types and thus provides a representative overview of estimation performance across different components of the model. 
We highlight the scenario with normally distributed errors and high reliability ($0.8$), as this scenario yielded the highest convergence rates across all estimation methods. 
Complete tabular results for all scenarios are provided in the appendix.

Figure~\ref{fig-perf-twofac} summarises the performance of our proposed eRBM and iRBM estimators relative to ML for the two-factor model, reporting mean bias, RMSE, PU, and coverage rates of Wald-type confidence intervals. 
Both eRBM and iRBM show clear reductions in bias compared to ML, with corresponding RMSE that is smaller or at least comparable, indicating that the bias reduction is not achieved at the expense of increased variability. 
PU for both methods are closer to the nominal $0.5$, reflecting reduced median bias, and the centred distributions of the eRBM and iRBM estimates are visibly shifted toward zero relative to ML (see Appendix Figure~\ref{fig-centdist-twofac}). 
As expected, the variability of all estimators decreases with increasing $n$. 
Finally, Wald-type confidence intervals based on eRBM and iRBM estimates achieve coverage rates that are closer to the nominal level than those based on ML. 
In small-sample scenarios, our methods improved coverage by as much as five percentage points relative to ML.
It is worth noting, however, that performance for $n = 15$---particularly under non-normal errors---is more difficult to assess due to the limited number of samples yielding acceptable estimates.

Figures~\ref{fig:twofac-bias} and \ref{fig:twofac-rmse} present the relative mean bias and RMSE, respectively, for the two-factor SEM across all estimation methods, based on samples yielding acceptable estimates. 
These figures allow for a direct comparison with the results reported in \textcite{dhaene2022resampling}. 
The bias performance of eRBM and iRBM was broadly comparable to the established resampling-based methods. 
However, the resampling-based methods tended to achieve slightly lower bias, particularly for variance parameters in the presence of non-normal errors, but this came at the cost of greater variability. 
In contrast, eRBM and iRBM did not exhibit this trade-off, delivering stable performance across most scenarios. 
A clear exception arises in the low-reliability, small-sample settings ($n = 15, 20$), where eRBM experienced estimation difficulties that led to rejection of roughly $60\%$ of parameter estimates. 
This resulted in inflated mean bias estimates for those scenarios. 

\subsection{Growth Curve Model: Bias, RMSE, and Coverage}\label{growth-curve-model-results}

The latent GCM exhibited generally higher rates of successful estimation than the two-factor SEM, even in the challenging small-sample, low-reliability settings.
To illustrate performance under difficult conditions, we focus on the scenario with normally distributed errors and low reliability ($0.5$). 
Of the six estimable parameters in this model, we report results for 
the residual variance $\theta_{11}$, the latent variances $\psi_{11}$ and $\psi_{22}$, and the latent covariance $\psi_{12}$.
For brevity, we do not present results for the latent intercept parameters. 
Because the item intercepts and selected loadings are fixed for identification, the latent intercepts are directly estimated from the data rather than inferred as latent quantities (as suggested by \eqref{eq-marg-mean} and \eqref{eq-scores-gcm-alpha}), resulting in negligible bias across all methods and conditions. 
Our discussion therefore focuses on the variance and covariance parameters, which are more challenging to estimate and provide a more meaningful assessment of bias-reduction performance.

\begin{figure}[htbp]
\centering
\includegraphics[width=1\linewidth,keepaspectratio]{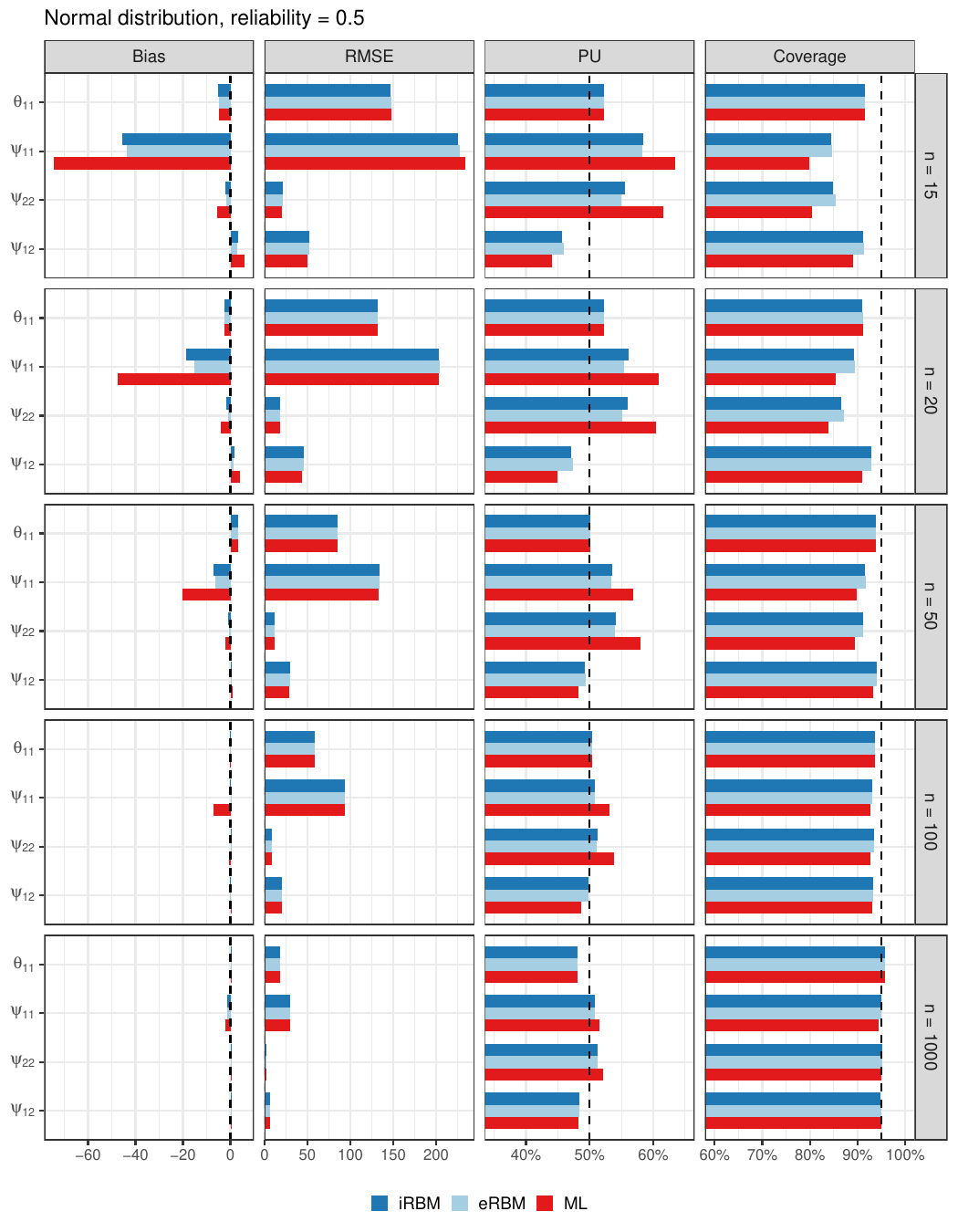}
\caption{\label{fig-perf-growth}Performance metrics (bias, RMSE, probability of underestimation, and coverage) of the ML, eRBM, and iRBM estimators for the latent GCM. Vertical dashed lines indicate the ideal values for each metric.}
\end{figure}%

\begin{figure}[p]
\centering

\begin{subfigure}{\linewidth}
  \centering
  \includegraphics[width=1\linewidth,keepaspectratio]{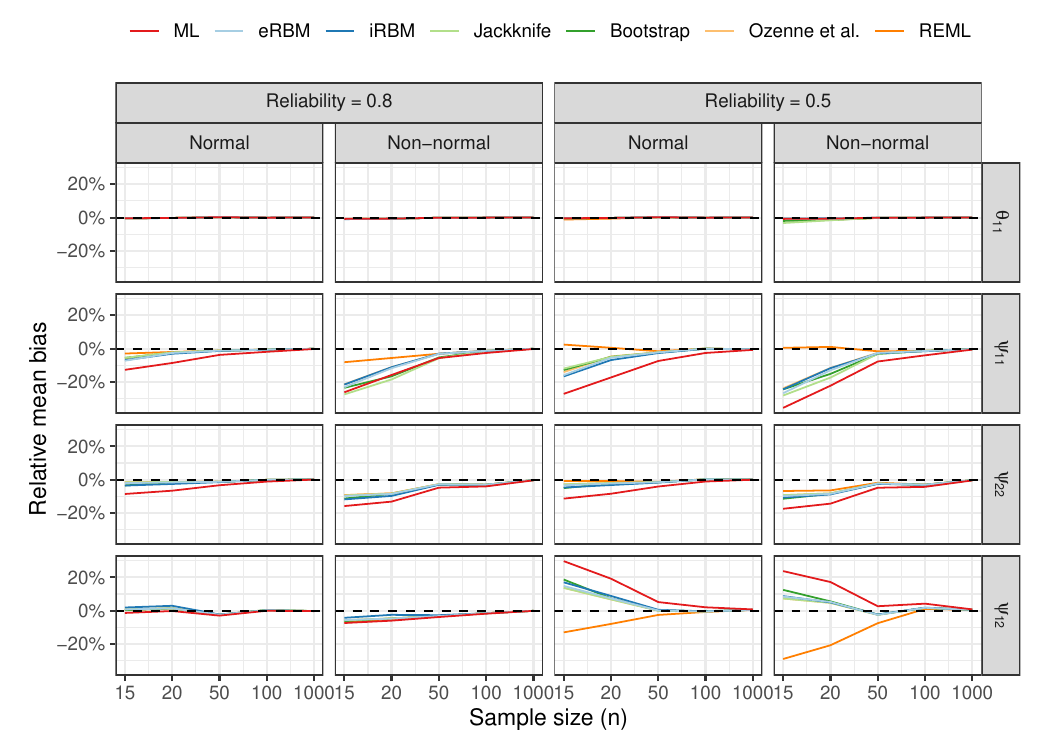}
  \caption{Relative mean bias.}
  \label{fig-meanbias-growth}
\end{subfigure}

\vspace{1em} 

\begin{subfigure}{\linewidth}
  \centering
\includegraphics[width=1\linewidth,keepaspectratio]{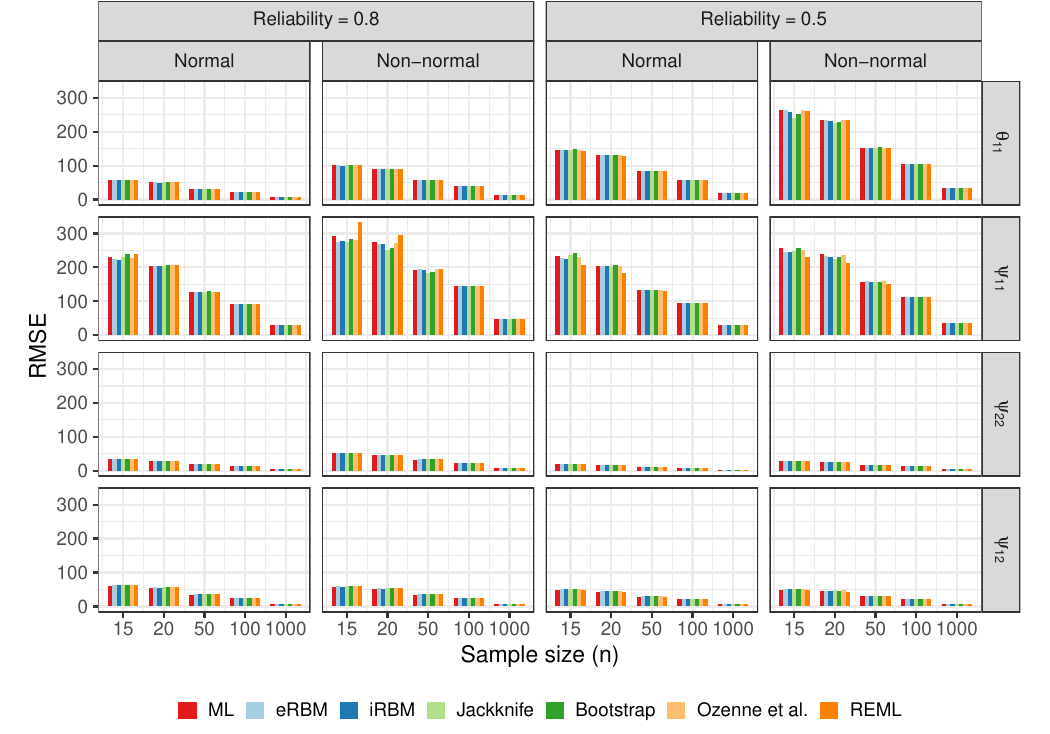}
  \caption{Root mean squared error (RMSE).}
  \label{fig-rmse-growth}
\end{subfigure}

\caption{\label{fig-growth}
Comparison of the ML, eRBM, and iRBM estimators with jackknife, bootstrap, and Ozenne corrections for estimation of the latent GCM.}
\end{figure}

Figure~\ref{fig-perf-growth} summarises the performance of eRBM and iRBM relative to ML for this model. 
Both methods produced clear bias reduction across all parameters, with RMSE values remaining comparable to those of ML, indicating that the bias reduction was not achieved at the cost of inflated variability. 
Among the parameters considered, the largest gains were observed for the latent variance $\psi_{11}$, whose true value was set to $\psi_{11} = 275$. 
In the $n = 15$ setting, bias was reduced by up to 30 units (approximately $10.9\%$). 

Median bias was also reduced, as reflected in probabilities of underestimation closer to the nominal $0.5$, whereas ML estimates tended to systematically overshoot. 
As with the two-factor SEM, the centred distributions of eRBM and iRBM estimates (Appendix Figure~\ref{fig-centdist-growth}) are shifted closer to zero than those of ML, showcasing their improved calibration. 
Wald-type confidence intervals constructed from eRBM and iRBM estimates exhibited coverage rates that were similar to or better than those based on ML, particularly for small to moderate sample sizes.
Compared to ML, these results confirm that both eRBM and iRBM deliver substantial improvements in bias and coverage for the latent GCM while maintaining stable variability across sample sizes.

We next compared eRBM and iRBM with other bias-reduction methods, including REML, jackknife, bootstrap, and the analytic approach of \textcite{ozenne2020small}. 
Figures~\ref{fig-meanbias-growth} and \ref{fig-rmse-growth} display the relative mean bias and RMSE, respectively, for the latent GCM across all methods based on samples with acceptable estimates. 
A striking feature is the very small bias observed for $\theta_{11}$, which likely reflects the stabilizing effect of constraining all residual variances to be equal across measurement occasions.

Under normally distributed data, REML generally achieved the lowest relative mean bias, consistent with its theoretical optimality for variance component estimation under normality. 
However, in low-reliability scenarios, the REML estimator exhibited considerable bias for the latent slope–intercept covariance $\psi_{12}$ in small samples. 
Moreover, REML’s performance deteriorated noticeably under non-normal data conditions, highlighting its lack of robustness to departures from normality.

By contrast, eRBM and iRBM maintained stable performance across all conditions. 
Their bias performance was consistently comparable---and in some cases superior---to that of the resampling-based methods and the Ozenne correction, without exhibiting the strong sensitivity to distributional assumptions observed for REML. 
In low-reliability scenarios, our methods substantially reduced the bias of $\psi_{12}$ relative to both ML and REML, particularly in small samples where this parameter was most unstable.


\section{Discussion}\label{discussion}

This paper has demonstrated the application of the empirical bias-reduction methods of \textcite{kosmidis2024empirical} to SEMs. 
RBM estimation shares the same spirit as \citeauthor{firth1993bias}'s (\citeyear{firth1993bias}) adjusted-score approach in that both aim to remove the $O(n^{-1})$ term in the bias expansion of the MLE. 
However, RBM achieves this by constructing an adjustment that is guaranteed to be the gradient of a penalty function, thus ensuring a penalized-likelihood interpretation. The resulting estimators are generally close to, but not identical with, those from the adjusted-score approach \autocite{kosmidis2024empirical}.
Both the iRBM and eRBM implementations are straightforward to apply within the SEM framework and require only the log-likelihood, its gradient, and Hessian. 
Our simulation study illustrates that these approaches are practically feasible and broadly effective for bias reduction in a range of SEM settings.

Across both the two-factor SEM and latent GCM, iRBM and eRBM consistently reduced finite-sample bias relative to ML estimation. 
These gains were most pronounced in small to moderate samples, where ML estimates are known to suffer from substantial bias. 
In addition, the methods improved the coverage probabilities of Wald-type confidence intervals, bringing them closer to their nominal levels, and did so without inflating estimator variability. 
Notably, the methods also maintained robustness under non-normal data distributions, particularly for the latent GCM, where REML and ML-based approaches showed noticeable degradation.

Convergence rates were uniformly high for $n \geq 50$, and iRBM in particular maintained near-perfect acceptance rates even in the most challenging scenarios. 
By contrast, eRBM showed reduced acceptance for $n = 15$ and $n = 20$ in the two-factor SEM under low reliability, with roughly $60\%$ of estimates being rejected due to non-positive definite covariance matrices or unstable standard errors. 
This pattern reflects a known limitation of post hoc corrections: 
the adjusted estimates can be pushed outside the parameter space, leading to infeasible solutions. 
This challenge is shared by other post hoc methods such as bootstrap, which also exhibited very low acceptance rates in the smallest samples. 
These results suggest that caution is warranted when applying post hoc bias corrections in highly constrained or small-sample settings.

Compared to traditional resampling-based approaches such as jackknife and bootstrap, iRBM and eRBM achieved a more favourable bias–variance trade-off. 
While resampling sometimes produced slightly lower bias, this was often accompanied by greater variability in the estimates, and in the case of bootstrap, a much higher computational burden. 
In contrast, iRBM and eRBM deliver bias reduction at a fraction of the computational cost, requiring only a single optimisation rather than repeated refitting across hundreds of resamples. 
For a modest sample size of $n=100$, eRBM required only $0.605$ seconds per replication on average for the two-factor SEM (c.f. bootstrap $53.2$ seconds; jackknife $9.77$ seconds), and $0.317$ seconds for the latent GCM (c.f. bootstrap $42.8$ seconds; jackknife $8.56$ seconds), representing significant reductions in runtime\footnote{Runtimes were measured on a MacBook Air M2 (8-core CPU @ 3.50 GHz, 16 GB RAM).} compared to resampling methods.
Although iRBM was slower than eRBM due to the penalized likelihood optimisation (averaging $13.0$ seconds per replication for the two-factor SEM and $1.20$ seconds for the latent GCM), it was still substantially faster than bootstrap while yielding comparable or superior statistical performance. 
REML, while theoretically optimal under normality, performed poorly under non-normal data, underscoring the importance of robust approaches such as RBM that do not rely heavily on distributional assumptions.

From a computational perspective, our implementation uses analytic casewise score vectors (Appendix \ref{sec-derivatives}), while the Hessian and its inverse are obtained numerically at the current $\vartheta$.
For iRBM, each evaluation of the penalized log-likelihood requires computing the inverse Hessian and the cross-product of the score contributions at the current $\vartheta$.
This repeated numerical differentiation at every optimisation step is the primary contributor to the higher runtime of iRBM. 
By comparison, eRBM performs the bias correction post hoc, which avoids repeated evaluations inside the optimiser and is therefore somewhat faster for SEMs. 
eRBM would be useful in settings where a model has already been fitted and a quick, post-hoc adjustment is desired.
Nevertheless, our recommendation---reflected in the default of the \texttt{\{brlavaan\}} package---is to use iRBM, as it performs the correction implicitly and is less prone to producing infeasible estimates.
Future computational improvements could include analytic derivations of the $A(\vartheta)$ matrix for common SEM specifications, as well as more efficient matrix factorization and quasi-Newton updates to reduce the cost of recomputing the inverse Hessian. 
Such improvements could dramatically reduce runtimes, making iRBM competitive with eRBM while preserving its robustness advantages.

Relatedly, a computational trade-off presents itself as models become more complex.
For instance, when more than two factors are involved and the dimension of the parameter vector $\vartheta$ expands significantly. 
Introducing additional latent variables necessitates estimating more factor loadings and a larger latent covariance matrix, which increases the size of the Hessian and variability matrix. 
Because of the aforementioned numerical routine involved, the runtime for iRBM is expected to increase non-linearly with the number of parameters. 
Then, the statistical arguments for using iRBM become even stronger in these high-dimensional settings. 
It is well-documented that as model complexity grows relative to sample size, the finite-sample properties of ML estimators deteriorate \parencite{jackson2003revisiting,wolf2013sample}. 
This `model size effect' typically manifests as increased bias in variance components \parencite{smid2020bayesian} and factor correlations \parencite{gerbing1985effects}, particularly when measurement quality is low or loadings are weakly identified \parencite{maccallum1999sample}. 
Furthermore, the standard ML estimator becomes increasingly prone to non-convergence or inadmissible solutions (e.g., Heywood cases), which can compromise the feasibility of the post-hoc eRBM correction in practice. 
On the other hand, iRBM operates on a penalized likelihood surface that often exhibits improved numerical behaviour and mitigates instability. 
Consequently, while iRBM may require more computational time per replication in complex models, it effectively serves as a self-regularizing procedure that is less dependent on the existence of a stable initial ML solution, supporting our recommendation of iRBM as the preferred option for small-sample SEM.

Returning to potential future work, another direction is to explore plugin penalties \parencite[e.g.][]{sterzinger2023maximum} to keep iRBM estimates within the interior of the parameter space. 
Such penalties could improve numerical stability, limit exploration of ill-conditioned regions, and potentially shorten runtimes. 
Although the proportion of unacceptable estimates was small, we did observe occasional iRBM estimation failures for $n \leq 50$ in both the two-factor SEM and latent GCM (Table~\ref{tbl-convsucc}). 
Light regularization of this type could therefore further enhance the stability of iRBM in these challenging small-sample scenarios where the likelihood surface is nearly flat.
Extensions to more complex SEMs---such as mediation models, models with latent interactions, and those estimated using alternatives to ML (e.g., WLS, DWLS)---are also a natural next step, given the generality of the M-estimation framework underlying RBM. 
Further investigation into non-normal likelihoods and robust estimation approaches would broaden the reach of RBM-based bias reduction to a wider class of psychometric and structural modelling problems.

\section{Conclusion}

Taken together, our results provide strong evidence that RBM-based bias reduction offers a practical and statistically sound alternative to both ML and resampling-based approaches for SEMs. 
The proposed iRBM and eRBM estimators consistently reduced small-sample bias, improved confidence interval coverage, and maintained robustness under non-normal data, all while being substantially more computationally efficient than traditional resampling methods. 
We believe these results make a compelling case for the routine use of RBM estimation in applied SEM analyses, particularly in studies with moderate to small sample sizes where bias can materially impact substantive conclusions.


\section*{Online Supplementary Materials}

All code, simulation results, and replication scripts for this study are  available at \url{https://osf.io/z8aky/}, and the latest version of the \texttt{\{brlavaan\}} package is available at the GitHub repo \texttt{haziqj/brlavaan} accessible at \url{https://github.com/haziqj/brlavaan}.

\printbibliography

\appendix

\section{Derivatives}
\label{sec-derivatives}

\subfile{derivatives}

\newpage
\section{Additional Plots}

\begin{figure}[H]
\centering
\includegraphics[width=1\linewidth,height=\textheight,keepaspectratio]{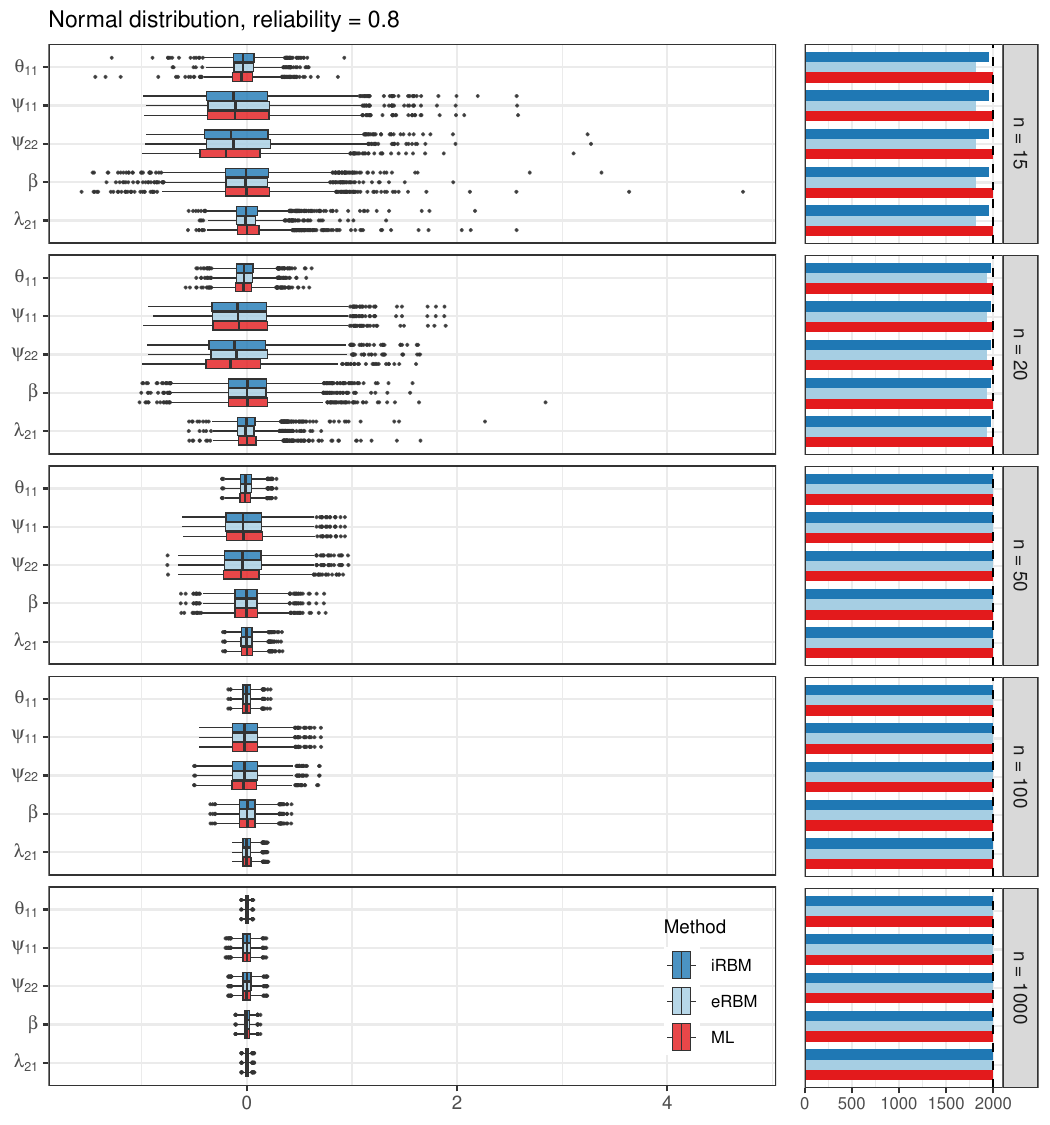}
\caption{\label{fig-centdist-twofac}Centred distributions of estimates for the two-factor SEM (left panel) using normally generated data at 80\% reliability. Unacceptable cases and extreme estimates have been excluded. The right panel displays the number of estimates used to compute the summary statistics.}
\end{figure}%

\begin{figure}[H]
\centering
\includegraphics[width=1\linewidth,height=\textheight,keepaspectratio]{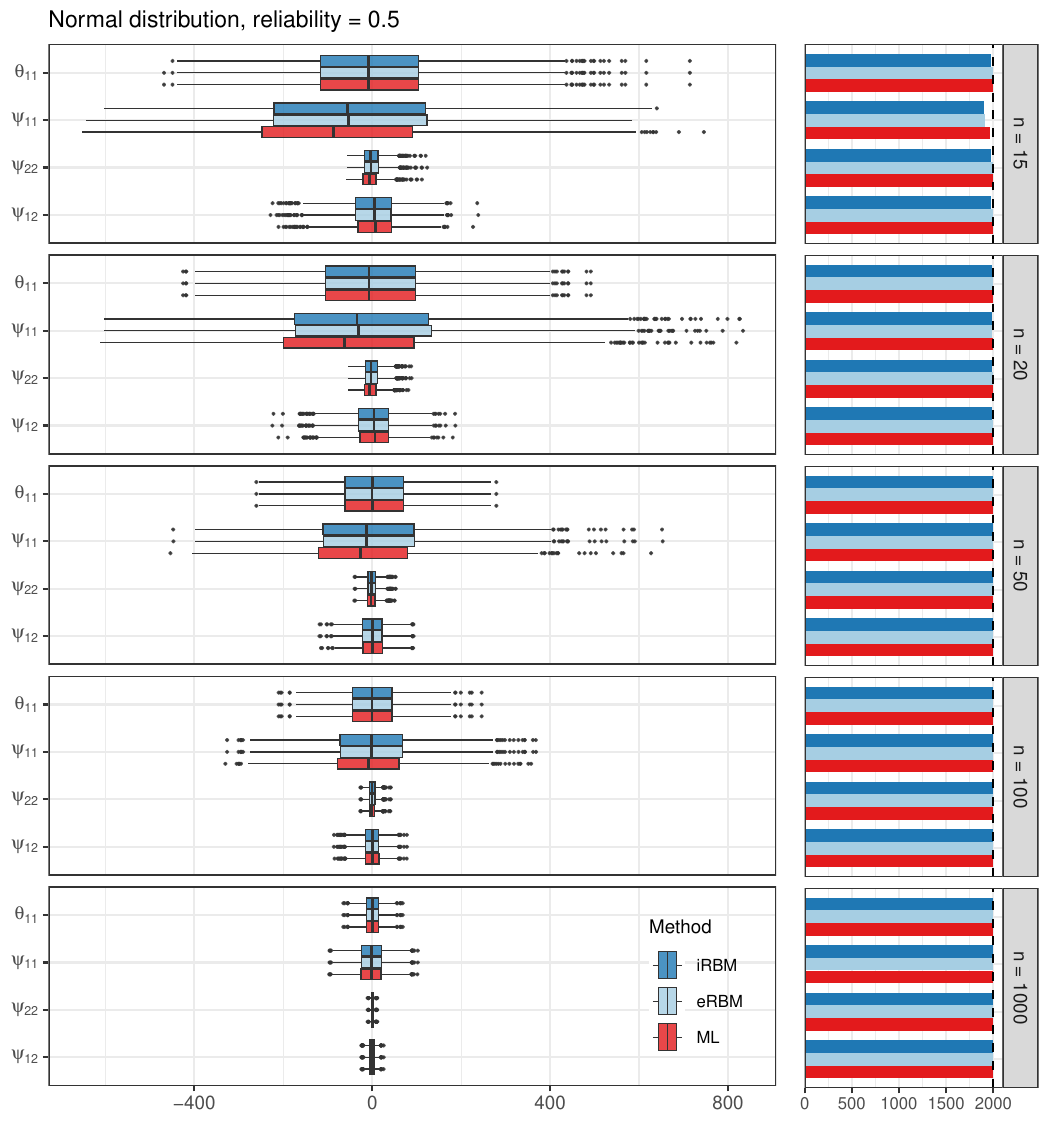}
\caption{\label{fig-centdist-growth}Centred distributions of estimates for the growth curve model (left panel) using normally generated data at 50\% reliability. Unacceptable estimates have been excluded. The right panel displays the number of estimates used to compute the summary statistics.}
\end{figure}%

\section{Tables}\label{b-tables}

\subfile{tables}

\end{document}

%% file: derivatives.tex
For the general SEM, we derive analytic expressions for the derivatives of \eqref{eq-sem_lik} with respect to each of the components $\vartheta_a$ of the parameter vector $\vartheta$.
With $\bar{\mathbf y}$ and $\mathbf S$ the sample mean vector and (biased) sample covariance matrix respectively, the derivatives of the log-likelihood with respect to $\boldsymbol\mu:=\boldsymbol\mu(\vartheta)$ and $\boldsymbol\Sigma:=\boldsymbol\Sigma(\vartheta)$ are
\begin{align}
\frac{\partial\ell}{\partial\boldsymbol\mu} &= n\boldsymbol\Sigma^{-1} (\bar{\mathbf y} - \boldsymbol\mu) =: n \bm\ell_\mu \label{eq-score-mu} \\
\frac{\partial\ell}{\partial\boldsymbol\Sigma} &= \frac{n}{2} (\boldsymbol\Sigma^{-1}\mathbf S \boldsymbol\Sigma^{-1} - \boldsymbol\Sigma^{-1}) =: \frac{n}{2} \mathbf L_\Sigma,
\end{align}
noting that $\bm\ell_\mu$ is a $p$-vector while $\mathbf L_\Sigma$ is a $p \times p$ symmetric matrix.
The chain rule \autocite[Sec. 2.8.1,][]{petersen2012matrix} gives us
\begin{equation}
\frac{\partial\ell}{\partial\vartheta_a} 
= n \bm\ell_\mu^\top \frac{\partial\boldsymbol\mu}{\partial\vartheta_a} + \frac{n}{2} \tr\left(\mathbf L_\Sigma \frac{\partial\boldsymbol\Sigma}{\partial\vartheta_a} \right).
\end{equation}

For a scalar entry $m_{ij}$ of the matrix $\mathbf M \in \mathbb R^{p\times q}$, the matrix defined by
$$
\mathbf R^{ij} = \frac{\partial \mathbf M}{\partial m_{ij}} = \begin{cases}
    \mathbf J^{ij}  &\mathbf M \text{ has no special structure} \\
    \mathbf J^{ij} + \mathbf J^{ji} - \mathbf J^{ij}\mathbf J^{ij} &\mathbf M \text{ is symmetric} 
\end{cases}
$$
is referred to as the \emph{structure matrix} of $\mathbf M$ \autocite[Eq. 134,][]{petersen2012matrix}.
Here, $\mathbf J^{ij} \in \mathbb R^{p \times q}$ is the single-entry matrix with a `$1$' at the $(i,j)$ position and 0 elsewhere.
Write
\begin{equation}
\tilde{\mathbf B} := (\mathbf I-\mathbf B)^{-1},
\qquad 
\boldsymbol\kappa := \tilde{\mathbf B} \boldsymbol\alpha,
\qquad\text{and}\qquad
\tilde{\boldsymbol\Psi} := \tilde{\mathbf B} \boldsymbol\Psi\tilde{\mathbf B}^\top,
\end{equation}
so that \eqref{eq-marg-mean} and \eqref{eq-marg-var} becomes
\begin{align}
\boldsymbol\mu := \boldsymbol\mu(\vartheta) &= \boldsymbol\nu+\boldsymbol\Lambda\boldsymbol\kappa \\
\boldsymbol\Sigma := \boldsymbol\Sigma(\vartheta) &= \boldsymbol\Lambda\tilde{\boldsymbol\Psi}\boldsymbol\Lambda^\top+\boldsymbol\Theta.
\end{align}

For the mean parameter, we have the differentials
\begin{equation}
d\boldsymbol\mu 
= d\boldsymbol\nu
+(d\boldsymbol\Lambda) \boldsymbol\kappa
+\boldsymbol\Lambda\tilde{\mathbf B} (d\boldsymbol\alpha)
+\boldsymbol\Lambda\tilde{\mathbf B} (d\mathbf B) \boldsymbol\kappa,
\end{equation}
using the fact that $d\tilde{\mathbf B} = \tilde{\mathbf B} (d\mathbf B) \tilde{\mathbf B}$ \autocite[Eq. 59][]{petersen2012matrix}. 
Furthermore,
\begin{align}
\frac{\partial\boldsymbol\mu}{\partial \boldsymbol\nu} &= \mathbf I & \frac{\partial\boldsymbol\mu}{\partial \lambda_{ij}} &= \mathbf R^{ij}\boldsymbol\kappa \\
\frac{\partial\boldsymbol\mu}{\partial \boldsymbol\alpha} &= \boldsymbol\Lambda \tilde{\mathbf B} & \frac{\partial\boldsymbol\mu}{\partial \beta_{kl}} &=  \boldsymbol\Lambda\tilde{\mathbf B} \mathbf R^{kl} \boldsymbol\kappa
\end{align}
and $\partial\boldsymbol\mu/\partial\boldsymbol\Psi = \mathbf 0$ and $\partial\boldsymbol\mu/\partial\boldsymbol\Theta = \mathbf 0$.

For the covariance parameter, we have the differentials
\begin{align*}
d\tilde{\boldsymbol\Psi} 
&= \tilde{\mathbf B}(d\boldsymbol\Psi)\tilde{\mathbf B}^\top + 
\tilde{\boldsymbol\Psi}(d\mathbf B)^\top \tilde{\mathbf B}^\top +
\tilde{\mathbf B}(d\mathbf B)\tilde{\boldsymbol\Psi} \\
d\boldsymbol\Sigma 
&= \boldsymbol\Lambda(d\tilde{\boldsymbol\Psi})\boldsymbol\Lambda^\top + 
\boldsymbol\Lambda\tilde{\boldsymbol\Psi}(d\boldsymbol\Lambda)^\top + 
(d\boldsymbol\Lambda)\tilde{\boldsymbol\Psi}\boldsymbol\Lambda^\top + 
d\boldsymbol\Theta.
\end{align*}
Also,
\begin{align}
\frac{\partial\boldsymbol\Sigma}{\partial \theta_{ij}} &= \mathbf R^{ij}
& \frac{\partial\boldsymbol\Sigma}{\partial \psi_{kl}} &= \boldsymbol\Lambda \tilde{\mathbf B} \mathbf R^{kl} (\boldsymbol\Lambda \tilde{\mathbf B})^\top \label{eq-diffSigmaa} \\
\frac{\partial\boldsymbol\Sigma}{\partial \lambda_{ij}} 
&= \boldsymbol\Lambda \tilde{\boldsymbol\Psi} (\mathbf R^{ij})^\top + \mathbf R^{ij} \tilde{\boldsymbol\Psi} \boldsymbol\Lambda^\top 
& \frac{\partial\boldsymbol\Sigma}{\partial \beta_{kl}} &=  \boldsymbol\Lambda\left( \tilde{\boldsymbol\Psi} (\mathbf R^{kl})^\top \tilde{\mathbf B}^\top + \tilde{\mathbf B} \mathbf R^{kl} \tilde{\boldsymbol\Psi} \right) \boldsymbol\Lambda^\top \label{eq-diffSigmab}
\end{align}
and $\partial\boldsymbol\Sigma/\partial\boldsymbol\nu = \mathbf 0$ and $\partial\boldsymbol\Sigma/\partial\boldsymbol\alpha = \mathbf 0$.

\subsection{Two-factor SEM}\label{a3-two-factor-model}

For the two-factor SEM that we considered, 
\begin{equation}
\tilde{\mathbf B} = \begin{bmatrix}1&0\\\beta&1\end{bmatrix},
\qquad 
\boldsymbol\kappa = \mathbf 0,
\qquad\text{and}\qquad
\tilde{\boldsymbol\Psi} = \begin{bmatrix}\psi_{11}&\beta\psi_{11}\\\beta\psi_{11}&\beta^2\psi_{11} + \psi_{22}\end{bmatrix},
\end{equation}
so all elements of $\vartheta$ appear only in $\boldsymbol\Sigma(\vartheta)$ and the relevant derivatives are only in \eqref{eq-diffSigmaa} and \eqref{eq-diffSigmab}.
Combining with the chain rule and simplifying gives us
\begin{align}
\frac{\partial \ell(\vartheta)}{\partial \lambda_{ij}} 
&= n\big(\mathbf L_\Sigma \boldsymbol\Lambda \tilde{\boldsymbol\Psi}\big)_{ij} &
\frac{\partial \ell(\vartheta)}{\partial \beta} 
&= n \left( \tilde{\mathbf B}^\top \boldsymbol\Lambda^\top \mathbf L_\Sigma \boldsymbol\Lambda \tilde{\boldsymbol\Psi} \right)_{12} \\
\frac{\partial \ell(\vartheta)}{\partial \theta_{ii}} 
&= \frac{n}{2} \big( \mathbf L_\Sigma \big)_{ii} &
\frac{\partial \ell(\vartheta)}{\partial \psi_{kk}} 
&= \frac{n}{2} \left( \tilde{\mathbf B}^\top \boldsymbol\Lambda^\top \mathbf L_\Sigma \boldsymbol\Lambda\tilde{\mathbf B} \right)_{kk} 
\end{align}

\subsection{Growth curve model}\label{a2-growth-curve-model}

For the latent GCM, $\tilde{\mathbf B} = \mathbf I$, $\boldsymbol\kappa = \boldsymbol\alpha$, and $\tilde{\boldsymbol\Psi} = \boldsymbol\Psi$.
The elements of $\boldsymbol\alpha$ only appear in $\boldsymbol\mu$, while the rest of the free parameters appear in $\boldsymbol\Sigma$.
Therefore,
\begin{align}
\frac{\partial \ell(\vartheta)}{\partial \boldsymbol\alpha} 
&= n \boldsymbol\Lambda^\top\boldsymbol\Sigma^{-1} (\bar{\mathbf y} - \boldsymbol\mu) &
\frac{\partial \ell(\vartheta)}{\partial \psi_{kl}} 
&= \frac{n}{2} \left( \boldsymbol\Lambda^\top \mathbf L_\Sigma \boldsymbol\Lambda \right)_{kl} \label{eq-scores-gcm-alpha} \\
\frac{\partial \ell(\vartheta)}{\partial \theta_{11}} 
&=  \frac{n}{2} \tr( \mathbf L_\Sigma)
\end{align}

%% file: tables.tex
\begin{sidewaystable}[htbp]
\centering

\tbl{\label{tbl-covr-twofac}Coverage rates 95\% confidence interval for the two-factor model.}

\fontsize{7.2pt}{7.7pt}\selectfont
\setlength{\tabcolsep}{2pt}  

\begin{tabular*}{0.8\linewidth}{@{\extracolsep{\fill}}l|rrrrrrrrrrrrrrrrrrrr}
\toprule
 & \multicolumn{10}{c}{Normal data} & \multicolumn{10}{c}{Non-Normal data} \\ 
\cmidrule(lr){2-11} \cmidrule(lr){12-21}
 & \multicolumn{5}{c}{Reliability = 0.8} & \multicolumn{5}{c}{Reliability = 0.5} & \multicolumn{5}{c}{Reliability = 0.8} & \multicolumn{5}{c}{Reliability = 0.5} \\ 
\cmidrule(lr){2-6} \cmidrule(lr){7-11} \cmidrule(lr){12-16} \cmidrule(lr){17-21}
 & 15 & 20 & 50 & 100 & 1000 & 15 & 20 & 50 & 100 & 1000 & 15 & 20 & 50 & 100 & 1000 & 15 & 20 & 50 & 100 & 1000 \\ 
\midrule\addlinespace[2.5pt]
\multicolumn{21}{l}{\(\theta_{11}\)} \\[2.5pt] 
\midrule\addlinespace[2.5pt]
ML & 0.82 & 0.87 & 0.92 & 0.94 & 0.95 & 0.86 & 0.90 & 0.94 & 0.95 & 0.95 & 0.74 & 0.77 & 0.83 & 0.89 & 0.94 & 0.77 & 0.80 & 0.86 & 0.90 & 0.94 \\ 
eRBM & 0.86 & 0.89 & 0.93 & 0.94 & 0.95 & 0.84 & 0.89 & 0.94 & 0.95 & 0.95 & 0.74 & 0.78 & 0.85 & 0.90 & 0.94 & 0.70 & 0.77 & 0.87 & 0.91 & 0.94 \\ 
iRBM & 0.85 & 0.89 & 0.93 & 0.94 & 0.95 & 0.83 & 0.87 & 0.93 & 0.95 & 0.95 & 0.75 & 0.77 & 0.85 & 0.90 & 0.94 & 0.71 & 0.75 & 0.86 & 0.91 & 0.94 \\ 
Ozenne et al. & 0.87 & 0.89 & 0.93 & 0.94 & 0.95 & 0.88 & 0.92 & 0.95 & 0.95 & 0.95 & 0.78 & 0.80 & 0.86 & 0.90 & 0.94 & 0.80 & 0.82 & 0.88 & 0.91 & 0.94 \\ 
Jackknife & 0.82 & 0.87 & 0.94 & 0.95 & 0.95 & 0.72 & 0.78 & 0.93 & 0.96 & 0.95 & 0.72 & 0.77 & 0.87 & 0.90 & 0.94 & 0.65 & 0.69 & 0.86 & 0.92 & 0.94 \\ 
Bootstrap & 0.75 & 0.81 & 0.91 & 0.94 & 0.94 & 0.69 & 0.75 & 0.90 & 0.95 & 0.95 & 0.64 & 0.69 & 0.83 & 0.89 & 0.94 & 0.57 & 0.62 & 0.79 & 0.89 & 0.94 \\ 
\midrule\addlinespace[2.5pt]
\multicolumn{21}{l}{\(\psi_{11}\)} \\[2.5pt] 
\midrule\addlinespace[2.5pt]
ML & 0.83 & 0.86 & 0.91 & 0.93 & 0.94 & 0.86 & 0.87 & 0.91 & 0.93 & 0.94 & 0.72 & 0.75 & 0.81 & 0.86 & 0.94 & 0.80 & 0.83 & 0.88 & 0.90 & 0.95 \\ 
eRBM & 0.78 & 0.81 & 0.89 & 0.92 & 0.94 & 0.84 & 0.80 & 0.76 & 0.84 & 0.94 & 0.71 & 0.72 & 0.79 & 0.85 & 0.94 & 0.84 & 0.83 & 0.79 & 0.81 & 0.94 \\ 
iRBM & 0.78 & 0.82 & 0.89 & 0.92 & 0.94 & 0.66 & 0.67 & 0.79 & 0.87 & 0.94 & 0.68 & 0.72 & 0.80 & 0.85 & 0.94 & 0.62 & 0.65 & 0.77 & 0.84 & 0.94 \\ 
Ozenne et al. & 0.90 & 0.91 & 0.93 & 0.94 & 0.94 & 0.85 & 0.90 & 0.96 & 0.94 & 0.94 & 0.80 & 0.82 & 0.84 & 0.87 & 0.94 & 0.82 & 0.87 & 0.93 & 0.92 & 0.95 \\ 
Jackknife & 0.81 & 0.84 & 0.92 & 0.93 & 0.94 & 0.64 & 0.71 & 0.89 & 0.93 & 0.94 & 0.70 & 0.73 & 0.82 & 0.86 & 0.94 & 0.58 & 0.64 & 0.84 & 0.90 & 0.94 \\ 
Bootstrap & 0.70 & 0.73 & 0.85 & 0.89 & 0.94 & 0.63 & 0.69 & 0.85 & 0.92 & 0.94 & 0.54 & 0.58 & 0.71 & 0.79 & 0.94 & 0.56 & 0.61 & 0.80 & 0.88 & 0.94 \\ 
\midrule\addlinespace[2.5pt]
\multicolumn{21}{l}{\(\psi_{22}\)} \\[2.5pt] 
\midrule\addlinespace[2.5pt]
ML & 0.78 & 0.81 & 0.88 & 0.92 & 0.94 & 0.78 & 0.82 & 0.89 & 0.93 & 0.94 & 0.68 & 0.71 & 0.80 & 0.86 & 0.93 & 0.75 & 0.78 & 0.85 & 0.89 & 0.94 \\ 
eRBM & 0.80 & 0.82 & 0.88 & 0.93 & 0.94 & 0.80 & 0.83 & 0.79 & 0.88 & 0.93 & 0.71 & 0.72 & 0.80 & 0.87 & 0.93 & 0.80 & 0.83 & 0.80 & 0.84 & 0.94 \\ 
iRBM & 0.79 & 0.81 & 0.88 & 0.93 & 0.94 & 0.65 & 0.70 & 0.81 & 0.90 & 0.93 & 0.68 & 0.71 & 0.80 & 0.87 & 0.93 & 0.62 & 0.67 & 0.80 & 0.86 & 0.94 \\ 
Ozenne et al. & 0.86 & 0.87 & 0.90 & 0.93 & 0.94 & 0.82 & 0.87 & 0.92 & 0.95 & 0.94 & 0.75 & 0.77 & 0.82 & 0.88 & 0.93 & 0.78 & 0.84 & 0.91 & 0.91 & 0.94 \\ 
Jackknife & 0.82 & 0.85 & 0.90 & 0.93 & 0.94 & 0.68 & 0.73 & 0.87 & 0.94 & 0.94 & 0.72 & 0.74 & 0.82 & 0.88 & 0.93 & 0.65 & 0.69 & 0.84 & 0.89 & 0.94 \\ 
Bootstrap & 0.72 & 0.76 & 0.85 & 0.91 & 0.94 & 0.66 & 0.70 & 0.85 & 0.93 & 0.94 & 0.59 & 0.63 & 0.74 & 0.82 & 0.93 & 0.59 & 0.64 & 0.80 & 0.88 & 0.94 \\ 
\midrule\addlinespace[2.5pt]
\multicolumn{21}{l}{\(\beta\)} \\[2.5pt] 
\midrule\addlinespace[2.5pt]
ML & 0.88 & 0.89 & 0.93 & 0.95 & 0.95 & 0.86 & 0.90 & 0.93 & 0.94 & 0.95 & 0.88 & 0.89 & 0.92 & 0.93 & 0.95 & 0.85 & 0.89 & 0.92 & 0.94 & 0.95 \\ 
eRBM & 0.87 & 0.88 & 0.93 & 0.95 & 0.95 & 0.86 & 0.90 & 0.93 & 0.94 & 0.95 & 0.85 & 0.86 & 0.91 & 0.93 & 0.95 & 0.79 & 0.85 & 0.92 & 0.93 & 0.95 \\ 
iRBM & 0.86 & 0.88 & 0.93 & 0.95 & 0.95 & 0.71 & 0.77 & 0.91 & 0.94 & 0.95 & 0.82 & 0.85 & 0.91 & 0.93 & 0.95 & 0.67 & 0.71 & 0.87 & 0.93 & 0.95 \\ 
Ozenne et al. & 0.89 & 0.90 & 0.93 & 0.95 & 0.95 & 0.84 & 0.89 & 0.94 & 0.95 & 0.95 & 0.89 & 0.90 & 0.92 & 0.93 & 0.95 & 0.83 & 0.87 & 0.92 & 0.94 & 0.95 \\ 
Jackknife & 0.93 & 0.93 & 0.95 & 0.95 & 0.95 & 0.86 & 0.88 & 0.93 & 0.95 & 0.95 & 0.92 & 0.93 & 0.94 & 0.94 & 0.95 & 0.82 & 0.85 & 0.92 & 0.95 & 0.96 \\ 
Bootstrap & 0.96 & 0.95 & 0.94 & 0.96 & 0.95 & 0.95 & 0.96 & 0.97 & 0.97 & 0.95 & 0.96 & 0.97 & 0.96 & 0.95 & 0.95 & 0.94 & 0.96 & 0.97 & 0.97 & 0.95 \\ 
\midrule\addlinespace[2.5pt]
\multicolumn{21}{l}{\(\lambda_{21}\)} \\[2.5pt] 
\midrule\addlinespace[2.5pt]
ML & 0.90 & 0.89 & 0.93 & 0.93 & 0.95 & 0.90 & 0.91 & 0.94 & 0.94 & 0.95 & 0.86 & 0.89 & 0.92 & 0.94 & 0.95 & 0.86 & 0.89 & 0.93 & 0.94 & 0.95 \\ 
eRBM & 0.87 & 0.86 & 0.93 & 0.93 & 0.95 & 0.84 & 0.82 & 0.89 & 0.91 & 0.95 & 0.80 & 0.84 & 0.91 & 0.93 & 0.95 & 0.77 & 0.84 & 0.88 & 0.91 & 0.94 \\ 
iRBM & 0.86 & 0.87 & 0.93 & 0.93 & 0.95 & 0.74 & 0.76 & 0.88 & 0.92 & 0.95 & 0.80 & 0.84 & 0.91 & 0.93 & 0.95 & 0.69 & 0.74 & 0.86 & 0.92 & 0.94 \\ 
Ozenne et al. & 0.91 & 0.90 & 0.94 & 0.94 & 0.95 & 0.86 & 0.88 & 0.94 & 0.94 & 0.95 & 0.86 & 0.90 & 0.92 & 0.94 & 0.95 & 0.82 & 0.87 & 0.92 & 0.95 & 0.95 \\ 
Jackknife & 0.91 & 0.92 & 0.94 & 0.94 & 0.95 & 0.78 & 0.82 & 0.92 & 0.95 & 0.95 & 0.88 & 0.92 & 0.94 & 0.95 & 0.95 & 0.77 & 0.81 & 0.90 & 0.95 & 0.95 \\ 
Bootstrap & 0.94 & 0.92 & 0.95 & 0.94 & 0.95 & 0.89 & 0.89 & 0.93 & 0.95 & 0.95 & 0.93 & 0.94 & 0.94 & 0.95 & 0.95 & 0.87 & 0.89 & 0.92 & 0.95 & 0.95 \\ 
\bottomrule
\end{tabular*}

\end{sidewaystable}%

\begin{sidewaystable}[htbp]
\tbl{\label{tbl-bias-twofac-80}Results two-factor model reliability 0.80.}
\centering

\fontsize{7pt}{7.7pt}\selectfont
\setlength{\tabcolsep}{2pt}  

\begin{tabular*}{\linewidth}{@{\extracolsep{\fill}}l|rrrrrrrrrrrrrrrrrrrrrrrrrrrrrr}
\toprule
 & \multicolumn{10}{c}{Relative mean bias} & \multicolumn{10}{c}{Relative median bias} & \multicolumn{10}{c}{Relative RMSE} \\ 
\cmidrule(lr){2-11} \cmidrule(lr){12-21} \cmidrule(lr){22-31}
 & \multicolumn{5}{c}{Normal data} & \multicolumn{5}{c}{Non-normal data} & \multicolumn{5}{c}{Normal data} & \multicolumn{5}{c}{Non-normal data} & \multicolumn{5}{c}{Normal data} & \multicolumn{5}{c}{Non-normal data} \\ 
\cmidrule(lr){2-6} \cmidrule(lr){7-11} \cmidrule(lr){12-16} \cmidrule(lr){17-21} \cmidrule(lr){22-26} \cmidrule(lr){27-31}
 & 15 & 20 & 50 & 100 & 1000 & 15 & 20 & 50 & 100 & 1000 & 15 & 20 & 50 & 100 & 1000 & 15 & 20 & 50 & 100 & 1000 & 15 & 20 & 50 & 100 & 1000 & 15 & 20 & 50 & 100 & 1000 \\ 
\midrule\addlinespace[2.5pt]
\multicolumn{31}{l}{\(\theta_{11}\)} \\[2.5pt] 
\midrule\addlinespace[2.5pt]
ML & -0.15 & -0.11 & -0.05 & -0.01 & 0.00 & -0.20 & -0.14 & -0.07 & -0.02 & -0.01 & -0.19 & -0.14 & -0.07 & -0.02 & 0.00 & -0.30 & -0.22 & -0.11 & -0.04 & -0.01 & 0.54 & 0.45 & 0.27 & 0.20 & 0.06 & 0.68 & 0.61 & 0.40 & 0.28 & 0.09 \\ 
eRBM & -0.10 & -0.08 & -0.03 & 0.00 & 0.00 & -0.21 & -0.13 & -0.05 & -0.01 & 0.00 & -0.15 & -0.11 & -0.05 & -0.01 & 0.00 & -0.31 & -0.21 & -0.10 & -0.03 & 0.00 & 0.52 & 0.44 & 0.27 & 0.20 & 0.06 & 0.64 & 0.58 & 0.40 & 0.29 & 0.09 \\ 
iRBM & -0.11 & -0.07 & -0.03 & 0.00 & 0.00 & -0.16 & -0.11 & -0.05 & -0.01 & 0.00 & -0.15 & -0.11 & -0.05 & -0.01 & 0.00 & -0.27 & -0.19 & -0.10 & -0.03 & 0.00 & 0.53 & 0.45 & 0.27 & 0.20 & 0.06 & 0.69 & 0.61 & 0.40 & 0.29 & 0.09 \\ 
Jackknife & 0.00 & 0.00 & -0.01 & 0.01 & 0.00 & -0.04 & -0.03 & -0.03 & 0.00 & 0.00 & -0.04 & -0.03 & -0.03 & 0.00 & 0.00 & -0.18 & -0.12 & -0.08 & -0.02 & 0.00 & 0.66 & 0.51 & 0.28 & 0.20 & 0.06 & 0.87 & 0.71 & 0.41 & 0.29 & 0.09 \\ 
Bootstrap & -0.04 & -0.02 & -0.01 & 0.01 & 0.00 & -0.10 & -0.06 & -0.04 & -0.01 & 0.00 & -0.07 & -0.05 & -0.03 & 0.00 & 0.00 & -0.20 & -0.14 & -0.08 & -0.02 & -0.01 & 0.59 & 0.50 & 0.28 & 0.20 & 0.06 & 0.73 & 0.65 & 0.42 & 0.29 & 0.09 \\ 
Ozenne et al. & -0.09 & -0.06 & -0.03 & 0.00 & 0.00 & -0.14 & -0.10 & -0.06 & -0.01 & 0.00 & -0.14 & -0.09 & -0.05 & -0.01 & 0.00 & -0.25 & -0.18 & -0.10 & -0.03 & -0.01 & 0.56 & 0.47 & 0.28 & 0.20 & 0.06 & 0.71 & 0.63 & 0.41 & 0.29 & 0.09 \\ 
\midrule\addlinespace[2.5pt]
\multicolumn{31}{l}{\(\psi_{11}\)} \\[2.5pt] 
\midrule\addlinespace[2.5pt]
ML & -0.08 & -0.05 & -0.02 & -0.02 & 0.00 & -0.09 & -0.09 & -0.04 & -0.02 & 0.00 & -0.11 & -0.08 & -0.03 & -0.02 & 0.00 & -0.20 & -0.17 & -0.07 & -0.04 & 0.00 & 0.40 & 0.35 & 0.22 & 0.16 & 0.05 & 0.57 & 0.50 & 0.35 & 0.26 & 0.08 \\ 
eRBM & -0.07 & -0.06 & -0.03 & -0.02 & 0.00 & -0.04 & -0.06 & -0.04 & -0.03 & 0.00 & -0.11 & -0.08 & -0.04 & -0.03 & 0.00 & -0.14 & -0.15 & -0.08 & -0.05 & 0.00 & 0.40 & 0.35 & 0.22 & 0.16 & 0.05 & 0.56 & 0.50 & 0.35 & 0.26 & 0.08 \\ 
iRBM & -0.09 & -0.06 & -0.03 & -0.02 & 0.00 & -0.09 & -0.09 & -0.04 & -0.02 & 0.00 & -0.12 & -0.09 & -0.04 & -0.02 & 0.00 & -0.19 & -0.18 & -0.08 & -0.05 & 0.00 & 0.40 & 0.35 & 0.22 & 0.16 & 0.05 & 0.57 & 0.51 & 0.35 & 0.26 & 0.08 \\ 
Jackknife & 0.01 & 0.00 & -0.01 & -0.01 & 0.00 & 0.01 & -0.01 & -0.02 & -0.02 & 0.00 & -0.04 & -0.03 & -0.02 & -0.02 & 0.00 & -0.13 & -0.11 & -0.06 & -0.04 & 0.00 & 0.47 & 0.39 & 0.22 & 0.16 & 0.05 & 0.70 & 0.58 & 0.36 & 0.26 & 0.08 \\ 
Bootstrap & 0.03 & 0.03 & 0.01 & 0.00 & 0.00 & 0.01 & 0.02 & 0.03 & 0.02 & 0.00 & -0.02 & 0.01 & 0.00 & -0.01 & 0.00 & -0.12 & -0.09 & -0.01 & -0.02 & 0.00 & 0.45 & 0.39 & 0.23 & 0.16 & 0.05 & 0.66 & 0.58 & 0.39 & 0.28 & 0.08 \\ 
Ozenne et al. & -0.01 & 0.00 & 0.00 & -0.01 & 0.00 & -0.03 & -0.04 & -0.02 & -0.01 & 0.00 & -0.05 & -0.03 & -0.01 & -0.01 & 0.00 & -0.15 & -0.13 & -0.05 & -0.03 & 0.00 & 0.42 & 0.36 & 0.22 & 0.16 & 0.05 & 0.60 & 0.52 & 0.35 & 0.26 & 0.08 \\ 
\midrule\addlinespace[2.5pt]
\multicolumn{31}{l}{\(\psi_{22}\)} \\[2.5pt] 
\midrule\addlinespace[2.5pt]
ML & -0.15 & -0.12 & -0.05 & -0.02 & 0.00 & -0.19 & -0.15 & -0.07 & -0.04 & 0.00 & -0.20 & -0.15 & -0.05 & -0.03 & 0.00 & -0.28 & -0.23 & -0.11 & -0.06 & 0.00 & 0.41 & 0.37 & 0.23 & 0.16 & 0.05 & 0.54 & 0.50 & 0.35 & 0.25 & 0.08 \\ 
eRBM & -0.07 & -0.07 & -0.03 & -0.01 & 0.00 & -0.09 & -0.09 & -0.05 & -0.03 & 0.00 & -0.13 & -0.10 & -0.04 & -0.02 & 0.00 & -0.19 & -0.17 & -0.09 & -0.05 & 0.00 & 0.41 & 0.37 & 0.23 & 0.16 & 0.05 & 0.54 & 0.50 & 0.35 & 0.25 & 0.08 \\ 
iRBM & -0.10 & -0.08 & -0.03 & -0.01 & 0.00 & -0.14 & -0.12 & -0.05 & -0.03 & 0.00 & -0.15 & -0.12 & -0.04 & -0.02 & 0.00 & -0.23 & -0.20 & -0.09 & -0.05 & 0.00 & 0.41 & 0.37 & 0.23 & 0.16 & 0.05 & 0.54 & 0.51 & 0.35 & 0.25 & 0.08 \\ 
Jackknife & 0.00 & -0.01 & -0.01 & 0.00 & 0.00 & -0.04 & -0.04 & -0.03 & -0.02 & 0.00 & -0.06 & -0.05 & -0.02 & -0.01 & 0.00 & -0.14 & -0.13 & -0.07 & -0.04 & 0.00 & 0.47 & 0.39 & 0.24 & 0.16 & 0.05 & 0.63 & 0.55 & 0.36 & 0.25 & 0.08 \\ 
Bootstrap & -0.01 & -0.01 & 0.00 & 0.00 & 0.00 & -0.05 & -0.02 & 0.01 & 0.00 & 0.00 & -0.07 & -0.05 & -0.01 & -0.01 & 0.00 & -0.16 & -0.13 & -0.04 & -0.03 & 0.00 & 0.45 & 0.39 & 0.24 & 0.16 & 0.05 & 0.61 & 0.56 & 0.38 & 0.27 & 0.08 \\ 
Ozenne et al. & -0.09 & -0.07 & -0.03 & -0.01 & 0.00 & -0.13 & -0.11 & -0.05 & -0.03 & 0.00 & -0.14 & -0.11 & -0.04 & -0.02 & 0.00 & -0.23 & -0.19 & -0.09 & -0.05 & 0.00 & 0.42 & 0.37 & 0.23 & 0.16 & 0.05 & 0.56 & 0.51 & 0.35 & 0.25 & 0.08 \\ 
\midrule\addlinespace[2.5pt]
\multicolumn{31}{l}{\(\beta\)} \\[2.5pt] 
\midrule\addlinespace[2.5pt]
ML & 0.03 & 0.05 & -0.02 & 0.01 & 0.00 & 0.03 & -0.04 & -0.02 & 0.01 & 0.00 & -0.01 & 0.02 & -0.02 & 0.01 & 0.00 & -0.06 & -0.11 & -0.05 & 0.00 & 0.00 & 1.21 & 1.02 & 0.57 & 0.40 & 0.12 & 1.32 & 1.07 & 0.59 & 0.40 & 0.12 \\ 
eRBM & 0.00 & 0.03 & -0.02 & 0.01 & 0.00 & -0.04 & -0.08 & -0.04 & 0.00 & 0.00 & -0.05 & 0.00 & -0.03 & 0.01 & 0.00 & -0.16 & -0.17 & -0.07 & -0.01 & 0.00 & 1.15 & 1.00 & 0.57 & 0.40 & 0.12 & 1.20 & 1.01 & 0.59 & 0.40 & 0.12 \\ 
iRBM & 0.00 & 0.04 & -0.02 & 0.01 & 0.00 & -0.04 & -0.08 & -0.03 & 0.00 & 0.00 & -0.04 & 0.02 & -0.03 & 0.01 & 0.00 & -0.15 & -0.15 & -0.07 & 0.00 & 0.00 & 1.17 & 1.01 & 0.57 & 0.40 & 0.12 & 1.26 & 1.03 & 0.59 & 0.40 & 0.12 \\ 
Jackknife & -0.06 & 0.00 & -0.03 & 0.01 & 0.00 & -0.09 & -0.12 & -0.05 & 0.00 & 0.00 & -0.09 & -0.02 & -0.03 & 0.01 & 0.00 & -0.23 & -0.25 & -0.08 & -0.01 & 0.00 & 1.29 & 1.03 & 0.57 & 0.40 & 0.12 & 1.71 & 1.24 & 0.59 & 0.40 & 0.12 \\ 
Bootstrap & -0.02 & 0.02 & -0.03 & 0.01 & 0.00 & -0.02 & -0.09 & -0.05 & -0.01 & 0.00 & -0.09 & -0.01 & -0.03 & 0.00 & 0.00 & -0.15 & -0.19 & -0.09 & -0.01 & 0.00 & 1.18 & 1.00 & 0.57 & 0.40 & 0.12 & 1.37 & 1.07 & 0.59 & 0.40 & 0.12 \\ 
Ozenne et al. & 0.03 & 0.05 & -0.02 & 0.01 & 0.00 & 0.03 & -0.04 & -0.02 & 0.01 & 0.00 & -0.01 & 0.02 & -0.02 & 0.01 & 0.00 & -0.06 & -0.11 & -0.05 & 0.00 & 0.00 & 1.21 & 1.02 & 0.57 & 0.40 & 0.12 & 1.33 & 1.07 & 0.59 & 0.40 & 0.12 \\ 
\midrule\addlinespace[2.5pt]
\multicolumn{31}{l}{\(\lambda_{21}\)} \\[2.5pt] 
\midrule\addlinespace[2.5pt]
ML & 0.02 & 0.01 & 0.00 & 0.00 & 0.00 & 0.01 & 0.01 & 0.00 & 0.00 & 0.00 & 0.00 & 0.00 & 0.00 & 0.00 & 0.00 & 0.00 & 0.00 & -0.01 & 0.00 & 0.00 & 0.21 & 0.18 & 0.10 & 0.07 & 0.02 & 0.23 & 0.19 & 0.11 & 0.07 & 0.02 \\ 
eRBM & -0.01 & -0.01 & 0.00 & 0.00 & 0.00 & -0.02 & -0.01 & -0.01 & 0.00 & 0.00 & -0.02 & -0.02 & -0.01 & 0.00 & 0.00 & -0.02 & -0.01 & -0.01 & 0.00 & 0.00 & 0.19 & 0.16 & 0.10 & 0.07 & 0.02 & 0.19 & 0.17 & 0.10 & 0.07 & 0.02 \\ 
iRBM & 0.01 & 0.00 & 0.00 & 0.00 & 0.00 & 0.00 & 0.00 & -0.01 & 0.00 & 0.00 & -0.01 & -0.01 & 0.00 & 0.00 & 0.00 & -0.01 & 0.00 & -0.01 & 0.00 & 0.00 & 0.21 & 0.17 & 0.10 & 0.07 & 0.02 & 0.23 & 0.19 & 0.11 & 0.07 & 0.02 \\ 
Jackknife & -0.02 & -0.02 & 0.00 & 0.00 & 0.00 & -0.03 & -0.02 & -0.01 & 0.00 & 0.00 & -0.03 & -0.02 & -0.01 & 0.00 & 0.00 & -0.04 & -0.02 & -0.02 & 0.00 & 0.00 & 0.22 & 0.17 & 0.10 & 0.07 & 0.02 & 0.31 & 0.22 & 0.11 & 0.07 & 0.02 \\ 
Bootstrap & -0.01 & -0.02 & -0.01 & 0.00 & 0.00 & -0.01 & -0.01 & -0.02 & 0.00 & 0.00 & -0.03 & -0.02 & -0.01 & 0.00 & 0.00 & -0.02 & -0.01 & -0.02 & 0.00 & 0.00 & 0.21 & 0.17 & 0.10 & 0.07 & 0.02 & 0.25 & 0.20 & 0.11 & 0.07 & 0.02 \\ 
Ozenne et al. & 0.02 & 0.01 & 0.00 & 0.00 & 0.00 & 0.01 & 0.01 & 0.00 & 0.00 & 0.00 & 0.00 & 0.00 & 0.00 & 0.00 & 0.00 & 0.00 & 0.00 & -0.01 & 0.00 & 0.00 & 0.21 & 0.18 & 0.10 & 0.07 & 0.02 & 0.23 & 0.19 & 0.11 & 0.07 & 0.02 \\ 
\bottomrule
\end{tabular*}

\end{sidewaystable}%

\begin{sidewaystable}[htbp]
\tbl{\label{tbl-bias-twofac-50}Results for two-factor model reliability 0.50.}
\centering

\fontsize{7pt}{7.7pt}\selectfont
\setlength{\tabcolsep}{2pt}  

\begin{tabular*}{\linewidth}{@{\extracolsep{\fill}}l|rrrrrrrrrrrrrrrrrrrrrrrrrrrrrr}
\toprule
 & \multicolumn{10}{c}{Relative mean bias} & \multicolumn{10}{c}{Relative median bias} & \multicolumn{10}{c}{Relative RMSE} \\ 
\cmidrule(lr){2-11} \cmidrule(lr){12-21} \cmidrule(lr){22-31}
 & \multicolumn{5}{c}{Normal data} & \multicolumn{5}{c}{Non-normal data} & \multicolumn{5}{c}{Normal data} & \multicolumn{5}{c}{Non-normal data} & \multicolumn{5}{c}{Normal data} & \multicolumn{5}{c}{Non-normal data} \\ 
\cmidrule(lr){2-6} \cmidrule(lr){7-11} \cmidrule(lr){12-16} \cmidrule(lr){17-21} \cmidrule(lr){22-26} \cmidrule(lr){27-31}
 & 15 & 20 & 50 & 100 & 1000 & 15 & 20 & 50 & 100 & 1000 & 15 & 20 & 50 & 100 & 1000 & 15 & 20 & 50 & 100 & 1000 & 15 & 20 & 50 & 100 & 1000 & 15 & 20 & 50 & 100 & 1000 \\ 
\midrule\addlinespace[2.5pt]
\multicolumn{31}{l}{\(\theta_{11}\)} \\[2.5pt] 
\midrule\addlinespace[2.5pt]
ML & -0.16 & -0.13 & -0.06 & -0.02 & 0.00 & -0.23 & -0.18 & -0.08 & -0.03 & -0.01 & -0.17 & -0.13 & -0.07 & -0.02 & 0.00 & -0.31 & -0.23 & -0.12 & -0.05 & -0.01 & 0.57 & 0.49 & 0.29 & 0.20 & 0.06 & 0.69 & 0.63 & 0.42 & 0.29 & 0.09 \\ 
eRBM & -0.21 & -0.15 & -0.03 & 0.00 & 0.00 & -0.32 & -0.26 & -0.07 & -0.01 & 0.00 & -0.24 & -0.16 & -0.04 & -0.01 & 0.00 & -0.41 & -0.30 & -0.10 & -0.03 & -0.01 & 0.49 & 0.41 & 0.27 & 0.20 & 0.06 & 0.60 & 0.53 & 0.38 & 0.28 & 0.09 \\ 
iRBM & -0.11 & -0.08 & -0.03 & 0.00 & 0.00 & -0.19 & -0.12 & -0.06 & -0.01 & 0.00 & -0.12 & -0.08 & -0.04 & -0.01 & 0.00 & -0.27 & -0.19 & -0.09 & -0.03 & -0.01 & 0.58 & 0.51 & 0.29 & 0.20 & 0.06 & 0.72 & 0.66 & 0.42 & 0.29 & 0.09 \\ 
Jackknife & 0.04 & 0.03 & 0.00 & 0.01 & 0.00 & 0.00 & 0.02 & -0.01 & 0.00 & 0.00 & 0.01 & 0.03 & -0.01 & 0.01 & 0.00 & -0.11 & -0.09 & -0.06 & -0.02 & 0.00 & 1.05 & 0.85 & 0.30 & 0.20 & 0.06 & 1.23 & 1.06 & 0.45 & 0.29 & 0.09 \\ 
Bootstrap & -0.06 & -0.03 & 0.00 & 0.01 & 0.00 & -0.13 & -0.07 & -0.02 & 0.01 & 0.00 & -0.05 & -0.01 & -0.01 & 0.00 & 0.00 & -0.22 & -0.14 & -0.06 & -0.02 & 0.00 & 0.68 & 0.60 & 0.31 & 0.20 & 0.06 & 0.79 & 0.74 & 0.47 & 0.30 & 0.09 \\ 
Ozenne et al. & -0.10 & -0.08 & -0.04 & -0.01 & 0.00 & -0.17 & -0.13 & -0.07 & -0.02 & -0.01 & -0.11 & -0.08 & -0.05 & -0.01 & 0.00 & -0.25 & -0.19 & -0.10 & -0.04 & -0.01 & 0.61 & 0.50 & 0.29 & 0.20 & 0.06 & 0.73 & 0.65 & 0.42 & 0.29 & 0.09 \\ 
\midrule\addlinespace[2.5pt]
\multicolumn{31}{l}{\(\psi_{11}\)} \\[2.5pt] 
\midrule\addlinespace[2.5pt]
ML & 0.02 & 0.01 & 0.00 & -0.01 & 0.00 & 0.06 & 0.04 & 0.02 & -0.01 & 0.00 & -0.08 & -0.07 & -0.03 & -0.03 & 0.00 & -0.11 & -0.09 & -0.03 & -0.04 & 0.00 & 0.66 & 0.60 & 0.39 & 0.27 & 0.08 & 0.79 & 0.70 & 0.49 & 0.34 & 0.11 \\ 
eRBM & 0.20 & 0.12 & -0.02 & -0.03 & 0.00 & 0.42 & 0.26 & 0.02 & -0.03 & 0.00 & 0.10 & 0.05 & -0.06 & -0.05 & 0.00 & 0.25 & 0.13 & -0.04 & -0.06 & -0.01 & 0.59 & 0.54 & 0.37 & 0.27 & 0.08 & 0.88 & 0.70 & 0.44 & 0.33 & 0.11 \\ 
iRBM & -0.01 & -0.02 & -0.03 & -0.02 & 0.00 & 0.02 & -0.01 & -0.01 & -0.03 & 0.00 & -0.12 & -0.11 & -0.06 & -0.04 & 0.00 & -0.15 & -0.13 & -0.06 & -0.06 & -0.01 & 0.68 & 0.63 & 0.39 & 0.27 & 0.08 & 0.81 & 0.73 & 0.49 & 0.34 & 0.11 \\ 
Jackknife & 0.02 & -0.02 & -0.02 & -0.02 & 0.00 & 0.05 & 0.01 & 0.00 & -0.02 & 0.00 & -0.19 & -0.17 & -0.06 & -0.04 & 0.00 & -0.25 & -0.19 & -0.07 & -0.06 & 0.00 & 1.34 & 1.07 & 0.42 & 0.27 & 0.08 & 1.47 & 1.21 & 0.55 & 0.34 & 0.11 \\ 
Bootstrap & 0.02 & 0.01 & -0.01 & -0.02 & 0.00 & 0.04 & 0.01 & 0.02 & -0.02 & 0.00 & -0.12 & -0.10 & -0.05 & -0.04 & 0.00 & -0.14 & -0.14 & -0.04 & -0.05 & 0.00 & 0.81 & 0.72 & 0.42 & 0.28 & 0.08 & 0.92 & 0.83 & 0.55 & 0.36 & 0.11 \\ 
Ozenne et al. & 0.09 & 0.06 & 0.02 & 0.00 & 0.00 & 0.13 & 0.09 & 0.04 & 0.00 & 0.00 & -0.02 & -0.02 & -0.01 & -0.02 & 0.00 & -0.05 & -0.04 & -0.01 & -0.03 & 0.00 & 0.71 & 0.63 & 0.39 & 0.27 & 0.08 & 0.85 & 0.74 & 0.50 & 0.34 & 0.11 \\ 
\midrule\addlinespace[2.5pt]
\multicolumn{31}{l}{\(\psi_{22}\)} \\[2.5pt] 
\midrule\addlinespace[2.5pt]
ML & -0.13 & -0.12 & -0.04 & -0.02 & 0.00 & -0.14 & -0.11 & -0.05 & -0.03 & 0.00 & -0.25 & -0.18 & -0.07 & -0.04 & 0.00 & -0.30 & -0.21 & -0.10 & -0.06 & 0.00 & 0.65 & 0.57 & 0.38 & 0.26 & 0.08 & 0.74 & 0.66 & 0.46 & 0.32 & 0.10 \\ 
eRBM & 0.06 & 0.04 & -0.03 & -0.02 & 0.00 & 0.10 & 0.12 & -0.02 & -0.03 & 0.00 & -0.01 & -0.04 & -0.06 & -0.04 & 0.00 & -0.06 & -0.02 & -0.07 & -0.06 & 0.00 & 0.62 & 0.52 & 0.38 & 0.26 & 0.08 & 0.71 & 0.65 & 0.44 & 0.32 & 0.10 \\ 
iRBM & -0.09 & -0.07 & -0.04 & -0.02 & 0.00 & -0.08 & -0.07 & -0.05 & -0.03 & 0.00 & -0.20 & -0.16 & -0.07 & -0.03 & 0.00 & -0.25 & -0.19 & -0.09 & -0.06 & 0.00 & 0.69 & 0.62 & 0.40 & 0.26 & 0.08 & 0.81 & 0.72 & 0.47 & 0.33 & 0.10 \\ 
Jackknife & 0.01 & -0.01 & -0.02 & -0.01 & 0.00 & -0.01 & -0.01 & -0.02 & -0.02 & 0.00 & -0.19 & -0.11 & -0.05 & -0.03 & 0.00 & -0.24 & -0.16 & -0.07 & -0.05 & 0.00 & 1.19 & 0.97 & 0.42 & 0.26 & 0.08 & 1.31 & 1.12 & 0.52 & 0.33 & 0.10 \\ 
Bootstrap & -0.04 & -0.03 & -0.01 & -0.01 & 0.00 & -0.07 & -0.03 & -0.01 & -0.02 & 0.00 & -0.17 & -0.11 & -0.05 & -0.02 & 0.00 & -0.24 & -0.15 & -0.07 & -0.04 & 0.00 & 0.76 & 0.69 & 0.42 & 0.27 & 0.08 & 0.86 & 0.78 & 0.51 & 0.34 & 0.10 \\ 
Ozenne et al. & -0.08 & -0.07 & -0.02 & -0.01 & 0.00 & -0.08 & -0.06 & -0.03 & -0.02 & 0.00 & -0.20 & -0.14 & -0.05 & -0.03 & 0.00 & -0.26 & -0.17 & -0.08 & -0.05 & 0.00 & 0.69 & 0.59 & 0.39 & 0.26 & 0.08 & 0.79 & 0.69 & 0.46 & 0.32 & 0.10 \\ 
\midrule\addlinespace[2.5pt]
\multicolumn{31}{l}{\(\beta\)} \\[2.5pt] 
\midrule\addlinespace[2.5pt]
ML & 0.09 & 0.14 & 0.00 & 0.03 & 0.00 & 0.16 & 0.05 & 0.00 & 0.01 & 0.01 & -0.09 & -0.03 & -0.05 & 0.01 & 0.00 & -0.11 & -0.13 & -0.06 & -0.02 & 0.00 & 1.79 & 1.54 & 0.79 & 0.53 & 0.16 & 1.95 & 1.55 & 0.82 & 0.55 & 0.16 \\ 
eRBM & 0.22 & 0.21 & -0.03 & 0.01 & -0.01 & 0.09 & 0.01 & -0.07 & -0.02 & 0.00 & 0.20 & 0.12 & -0.06 & -0.01 & 0.00 & 0.01 & -0.05 & -0.11 & -0.05 & 0.00 & 1.46 & 1.27 & 0.73 & 0.52 & 0.16 & 1.51 & 1.27 & 0.74 & 0.54 & 0.16 \\ 
iRBM & -0.01 & 0.02 & -0.03 & 0.02 & -0.01 & 0.02 & -0.08 & -0.05 & -0.01 & 0.00 & -0.22 & -0.16 & -0.08 & 0.00 & 0.00 & -0.30 & -0.33 & -0.11 & -0.04 & 0.00 & 1.61 & 1.43 & 0.78 & 0.52 & 0.16 & 1.84 & 1.49 & 0.82 & 0.55 & 0.16 \\ 
Jackknife & 0.01 & 0.06 & -0.07 & 0.00 & -0.01 & 0.22 & -0.15 & -0.09 & -0.03 & 0.00 & -0.24 & -0.15 & -0.09 & -0.02 & 0.00 & -0.19 & -0.35 & -0.16 & -0.06 & 0.00 & 4.06 & 2.65 & 0.82 & 0.52 & 0.16 & 4.68 & 3.12 & 0.92 & 0.54 & 0.16 \\ 
Bootstrap & 0.04 & 0.10 & -0.06 & 0.00 & -0.01 & 0.20 & -0.02 & -0.08 & -0.03 & 0.00 & -0.11 & -0.05 & -0.10 & -0.02 & 0.00 & -0.08 & -0.23 & -0.13 & -0.06 & 0.00 & 2.28 & 1.71 & 0.78 & 0.52 & 0.16 & 2.60 & 1.93 & 0.83 & 0.55 & 0.16 \\ 
Ozenne et al. & 0.08 & 0.13 & 0.00 & 0.03 & 0.00 & 0.14 & 0.05 & 0.00 & 0.01 & 0.01 & -0.10 & -0.04 & -0.05 & 0.01 & 0.00 & -0.12 & -0.14 & -0.06 & -0.02 & 0.00 & 1.79 & 1.55 & 0.79 & 0.53 & 0.16 & 1.95 & 1.57 & 0.82 & 0.55 & 0.16 \\ 
\midrule\addlinespace[2.5pt]
\multicolumn{31}{l}{\(\lambda_{21}\)} \\[2.5pt] 
\midrule\addlinespace[2.5pt]
ML & 0.05 & 0.05 & 0.04 & 0.01 & 0.00 & 0.01 & 0.03 & 0.01 & 0.02 & 0.00 & -0.03 & -0.02 & 0.00 & 0.00 & 0.00 & -0.05 & -0.03 & -0.02 & 0.00 & 0.00 & 0.53 & 0.47 & 0.28 & 0.18 & 0.06 & 0.55 & 0.47 & 0.29 & 0.19 & 0.06 \\ 
eRBM & -0.09 & -0.10 & -0.03 & -0.01 & 0.00 & -0.07 & -0.06 & -0.05 & -0.01 & 0.00 & -0.10 & -0.10 & -0.05 & -0.02 & 0.00 & -0.09 & -0.08 & -0.06 & -0.02 & -0.01 & 0.31 & 0.31 & 0.23 & 0.17 & 0.06 & 0.32 & 0.29 & 0.23 & 0.18 & 0.06 \\ 
iRBM & 0.08 & 0.05 & 0.02 & 0.00 & 0.00 & 0.07 & 0.04 & 0.00 & 0.01 & 0.00 & -0.03 & -0.03 & -0.01 & -0.01 & 0.00 & -0.05 & -0.03 & -0.03 & -0.01 & -0.01 & 0.64 & 0.55 & 0.28 & 0.18 & 0.06 & 0.69 & 0.58 & 0.30 & 0.19 & 0.06 \\ 
Jackknife & 0.03 & -0.01 & -0.04 & -0.01 & 0.00 & 0.10 & 0.03 & -0.06 & -0.01 & 0.00 & -0.14 & -0.12 & -0.06 & -0.02 & 0.00 & -0.10 & -0.08 & -0.07 & -0.02 & -0.01 & 1.37 & 0.89 & 0.26 & 0.17 & 0.06 & 1.53 & 1.15 & 0.31 & 0.18 & 0.06 \\ 
Bootstrap & 0.10 & 0.04 & -0.02 & -0.01 & 0.00 & 0.10 & 0.08 & -0.03 & -0.01 & 0.00 & -0.06 & -0.08 & -0.05 & -0.02 & 0.00 & -0.05 & -0.03 & -0.06 & -0.02 & -0.01 & 0.76 & 0.59 & 0.28 & 0.18 & 0.06 & 0.82 & 0.66 & 0.31 & 0.19 & 0.06 \\ 
Ozenne et al. & 0.05 & 0.05 & 0.04 & 0.01 & 0.00 & 0.02 & 0.04 & 0.01 & 0.02 & 0.00 & -0.02 & -0.02 & 0.00 & 0.00 & 0.00 & -0.05 & -0.02 & -0.02 & 0.00 & 0.00 & 0.55 & 0.48 & 0.28 & 0.18 & 0.06 & 0.56 & 0.48 & 0.29 & 0.19 & 0.06 \\ 
\bottomrule
\end{tabular*}

\end{sidewaystable}%

\begin{sidewaystable}[htbp]
\centering

\tbl{\label{tbl-covr-growth}Coverage rates of 95\% confidence interval for the latent growth curve model.}

\fontsize{7pt}{8.4pt}\selectfont
\setlength{\tabcolsep}{2pt}  

\begin{tabular*}{0.8\linewidth}{@{\extracolsep{\fill}}l|rrrrrrrrrrrrrrrrrrrr}
\toprule
 & \multicolumn{10}{c}{Normal data} & \multicolumn{10}{c}{Non-Normal data} \\ 
\cmidrule(lr){2-11} \cmidrule(lr){12-21}
 & \multicolumn{5}{c}{Reliability = 0.8} & \multicolumn{5}{c}{Reliability = 0.5} & \multicolumn{5}{c}{Reliability = 0.8} & \multicolumn{5}{c}{Reliability = 0.5} \\ 
\cmidrule(lr){2-6} \cmidrule(lr){7-11} \cmidrule(lr){12-16} \cmidrule(lr){17-21}
 & 15 & 20 & 50 & 100 & 1000 & 15 & 20 & 50 & 100 & 1000 & 15 & 20 & 50 & 100 & 1000 & 15 & 20 & 50 & 100 & 1000 \\ 
\midrule\addlinespace[2.5pt]
\multicolumn{21}{l}{\(\theta_{11}\)} \\[2.5pt] 
\midrule\addlinespace[2.5pt]
ML & 0.91 & 0.91 & 0.94 & 0.94 & 0.96 & 0.91 & 0.91 & 0.94 & 0.94 & 0.96 & 0.87 & 0.89 & 0.92 & 0.94 & 0.95 & 0.87 & 0.89 & 0.92 & 0.94 & 0.95 \\ 
eRBM & 0.91 & 0.91 & 0.94 & 0.94 & 0.96 & 0.91 & 0.91 & 0.94 & 0.94 & 0.96 & 0.87 & 0.89 & 0.92 & 0.94 & 0.95 & 0.87 & 0.89 & 0.92 & 0.94 & 0.95 \\ 
iRBM & 0.92 & 0.91 & 0.94 & 0.94 & 0.96 & 0.92 & 0.91 & 0.94 & 0.94 & 0.96 & 0.88 & 0.89 & 0.92 & 0.94 & 0.95 & 0.88 & 0.89 & 0.92 & 0.94 & 0.95 \\ 
Ozenne et al. & 0.92 & 0.92 & 0.94 & 0.94 & 0.96 & 0.92 & 0.92 & 0.94 & 0.94 & 0.96 & 0.88 & 0.90 & 0.92 & 0.94 & 0.95 & 0.88 & 0.90 & 0.92 & 0.94 & 0.95 \\ 
Jackknife & 0.92 & 0.92 & 0.94 & 0.94 & 0.96 & 0.92 & 0.92 & 0.94 & 0.94 & 0.96 & 0.88 & 0.90 & 0.92 & 0.94 & 0.95 & 0.88 & 0.90 & 0.92 & 0.94 & 0.95 \\ 
Bootstrap & 0.88 & 0.91 & 0.94 & 0.94 & 0.96 & 0.88 & 0.91 & 0.94 & 0.94 & 0.96 & 0.84 & 0.89 & 0.91 & 0.94 & 0.95 & 0.83 & 0.89 & 0.91 & 0.94 & 0.95 \\ 
REML & 0.67 & 0.70 & 0.83 & 0.93 & 1.00 & 0.27 & 0.29 & 0.41 & 0.59 & 0.99 & 0.41 & 0.41 & 0.54 & 0.69 & 0.99 & 0.14 & 0.15 & 0.23 & 0.34 & 0.83 \\ 
\midrule\addlinespace[2.5pt]
\multicolumn{21}{l}{\(\psi_{11}\)} \\[2.5pt] 
\midrule\addlinespace[2.5pt]
ML & 0.79 & 0.84 & 0.90 & 0.92 & 0.94 & 0.80 & 0.85 & 0.90 & 0.93 & 0.94 & 0.67 & 0.71 & 0.82 & 0.87 & 0.94 & 0.78 & 0.81 & 0.88 & 0.91 & 0.94 \\ 
eRBM & 0.84 & 0.88 & 0.92 & 0.93 & 0.95 & 0.85 & 0.89 & 0.92 & 0.93 & 0.95 & 0.71 & 0.75 & 0.84 & 0.88 & 0.94 & 0.82 & 0.84 & 0.90 & 0.92 & 0.94 \\ 
iRBM & 0.85 & 0.88 & 0.92 & 0.93 & 0.95 & 0.84 & 0.89 & 0.91 & 0.93 & 0.95 & 0.72 & 0.76 & 0.84 & 0.88 & 0.94 & 0.83 & 0.85 & 0.90 & 0.92 & 0.94 \\ 
Ozenne et al. & 0.85 & 0.89 & 0.92 & 0.93 & 0.95 & 0.85 & 0.90 & 0.92 & 0.93 & 0.95 & 0.73 & 0.76 & 0.84 & 0.88 & 0.94 & 0.83 & 0.85 & 0.90 & 0.92 & 0.94 \\ 
Jackknife & 0.83 & 0.88 & 0.92 & 0.93 & 0.95 & 0.84 & 0.89 & 0.91 & 0.93 & 0.95 & 0.69 & 0.73 & 0.83 & 0.88 & 0.94 & 0.80 & 0.83 & 0.90 & 0.92 & 0.94 \\ 
Bootstrap & 0.76 & 0.85 & 0.91 & 0.92 & 0.94 & 0.77 & 0.86 & 0.90 & 0.93 & 0.94 & 0.63 & 0.71 & 0.82 & 0.87 & 0.94 & 0.75 & 0.81 & 0.89 & 0.92 & 0.94 \\ 
REML & 0.58 & 0.66 & 0.87 & 0.96 & 1.00 & 0.52 & 0.61 & 0.81 & 0.93 & 1.00 & 0.42 & 0.47 & 0.68 & 0.82 & 1.00 & 0.46 & 0.51 & 0.73 & 0.88 & 1.00 \\ 
\midrule\addlinespace[2.5pt]
\multicolumn{21}{l}{\(\psi_{22}\)} \\[2.5pt] 
\midrule\addlinespace[2.5pt]
ML & 0.81 & 0.84 & 0.90 & 0.93 & 0.95 & 0.80 & 0.84 & 0.90 & 0.93 & 0.95 & 0.66 & 0.69 & 0.81 & 0.84 & 0.93 & 0.69 & 0.73 & 0.82 & 0.86 & 0.94 \\ 
eRBM & 0.87 & 0.88 & 0.91 & 0.93 & 0.95 & 0.85 & 0.87 & 0.91 & 0.93 & 0.95 & 0.70 & 0.73 & 0.82 & 0.86 & 0.93 & 0.75 & 0.77 & 0.84 & 0.87 & 0.94 \\ 
iRBM & 0.86 & 0.88 & 0.91 & 0.93 & 0.95 & 0.85 & 0.87 & 0.91 & 0.93 & 0.95 & 0.69 & 0.72 & 0.82 & 0.86 & 0.93 & 0.74 & 0.77 & 0.84 & 0.87 & 0.94 \\ 
Ozenne et al. & 0.88 & 0.89 & 0.91 & 0.93 & 0.95 & 0.86 & 0.88 & 0.91 & 0.93 & 0.95 & 0.71 & 0.74 & 0.82 & 0.86 & 0.93 & 0.76 & 0.78 & 0.84 & 0.87 & 0.94 \\ 
Jackknife & 0.86 & 0.88 & 0.91 & 0.93 & 0.95 & 0.85 & 0.87 & 0.91 & 0.93 & 0.95 & 0.70 & 0.73 & 0.82 & 0.85 & 0.93 & 0.74 & 0.77 & 0.84 & 0.87 & 0.94 \\ 
Bootstrap & 0.79 & 0.85 & 0.90 & 0.92 & 0.95 & 0.78 & 0.85 & 0.90 & 0.92 & 0.95 & 0.63 & 0.70 & 0.81 & 0.84 & 0.93 & 0.67 & 0.74 & 0.83 & 0.86 & 0.94 \\ 
REML & 0.99 & 1.00 & 1.00 & 1.00 & 1.00 & 0.95 & 0.96 & 0.99 & 1.00 & 1.00 & 0.95 & 0.97 & 1.00 & 1.00 & 1.00 & 0.90 & 0.92 & 0.98 & 1.00 & 1.00 \\ 
\midrule\addlinespace[2.5pt]
\multicolumn{21}{l}{\(\psi_{12}\)} \\[2.5pt] 
\midrule\addlinespace[2.5pt]
ML & 0.92 & 0.93 & 0.94 & 0.94 & 0.95 & 0.89 & 0.91 & 0.93 & 0.93 & 0.95 & 0.92 & 0.93 & 0.93 & 0.93 & 0.95 & 0.91 & 0.92 & 0.94 & 0.94 & 0.95 \\ 
eRBM & 0.94 & 0.94 & 0.95 & 0.94 & 0.95 & 0.91 & 0.93 & 0.94 & 0.93 & 0.95 & 0.94 & 0.94 & 0.94 & 0.94 & 0.95 & 0.93 & 0.93 & 0.95 & 0.94 & 0.95 \\ 
iRBM & 0.93 & 0.94 & 0.95 & 0.94 & 0.95 & 0.91 & 0.93 & 0.94 & 0.93 & 0.95 & 0.94 & 0.94 & 0.94 & 0.94 & 0.95 & 0.93 & 0.93 & 0.95 & 0.94 & 0.95 \\ 
Ozenne et al. & 0.94 & 0.95 & 0.95 & 0.94 & 0.95 & 0.92 & 0.94 & 0.94 & 0.93 & 0.95 & 0.95 & 0.94 & 0.94 & 0.94 & 0.95 & 0.94 & 0.94 & 0.95 & 0.95 & 0.95 \\ 
Jackknife & 0.93 & 0.93 & 0.94 & 0.94 & 0.95 & 0.91 & 0.92 & 0.94 & 0.93 & 0.95 & 0.93 & 0.93 & 0.94 & 0.94 & 0.95 & 0.92 & 0.93 & 0.95 & 0.94 & 0.95 \\ 
Bootstrap & 0.86 & 0.91 & 0.93 & 0.94 & 0.95 & 0.83 & 0.90 & 0.92 & 0.93 & 0.95 & 0.85 & 0.90 & 0.93 & 0.93 & 0.95 & 0.84 & 0.90 & 0.93 & 0.94 & 0.95 \\ 
REML & 0.63 & 0.65 & 0.79 & 0.89 & 1.00 & 0.71 & 0.73 & 0.88 & 0.96 & 1.00 & 0.64 & 0.67 & 0.80 & 0.89 & 1.00 & 0.70 & 0.74 & 0.87 & 0.95 & 1.00 \\ 
\bottomrule
\end{tabular*}

\end{sidewaystable}%

\begin{sidewaystable}[htbp]

\tbl{\label{tbl-bias-growth-80}Results growth curve model reliability 0.80.}
\centering

\fontsize{7pt}{8.4pt}\selectfont
\setlength{\tabcolsep}{2pt}  

\begin{tabular*}{\linewidth}{@{\extracolsep{\fill}}l|rrrrrrrrrrrrrrrrrrrrrrrrrrrrrr}
\toprule
 & \multicolumn{10}{c}{Relative mean bias} & \multicolumn{10}{c}{Relative median bias} & \multicolumn{10}{c}{Relative RMSE} \\ 
\cmidrule(lr){2-11} \cmidrule(lr){12-21} \cmidrule(lr){22-31}
 & \multicolumn{5}{c}{Normal data} & \multicolumn{5}{c}{Non-normal data} & \multicolumn{5}{c}{Normal data} & \multicolumn{5}{c}{Non-normal data} & \multicolumn{5}{c}{Normal data} & \multicolumn{5}{c}{Non-normal data} \\ 
\cmidrule(lr){2-6} \cmidrule(lr){7-11} \cmidrule(lr){12-16} \cmidrule(lr){17-21} \cmidrule(lr){22-26} \cmidrule(lr){27-31}
 & 15 & 20 & 50 & 100 & 1000 & 15 & 20 & 50 & 100 & 1000 & 15 & 20 & 50 & 100 & 1000 & 15 & 20 & 50 & 100 & 1000 & 15 & 20 & 50 & 100 & 1000 & 15 & 20 & 50 & 100 & 1000 \\ 
\midrule\addlinespace[2.5pt]
\multicolumn{31}{l}{\(\theta_{11}\)} \\[2.5pt] 
\midrule\addlinespace[2.5pt]
ML & 0.00 & 0.00 & 0.00 & 0.00 & 0.00 & -0.01 & -0.01 & 0.00 & 0.00 & 0.00 & -0.01 & -0.01 & 0.00 & 0.00 & 0.00 & -0.03 & -0.03 & 0.00 & 0.00 & 0.00 & 0.11 & 0.10 & 0.07 & 0.04 & 0.01 & 0.20 & 0.18 & 0.12 & 0.08 & 0.03 \\ 
eRBM & 0.00 & 0.00 & 0.00 & 0.00 & 0.00 & -0.01 & -0.01 & 0.00 & 0.00 & 0.00 & -0.01 & -0.01 & 0.00 & 0.00 & 0.00 & -0.03 & -0.03 & 0.00 & 0.00 & 0.00 & 0.11 & 0.10 & 0.07 & 0.04 & 0.01 & 0.20 & 0.18 & 0.12 & 0.08 & 0.03 \\ 
iRBM & 0.00 & 0.00 & 0.00 & 0.00 & 0.00 & -0.01 & -0.01 & 0.00 & 0.00 & 0.00 & -0.01 & 0.00 & 0.00 & 0.00 & 0.00 & -0.02 & -0.02 & 0.00 & 0.00 & 0.00 & 0.11 & 0.10 & 0.07 & 0.04 & 0.01 & 0.20 & 0.18 & 0.12 & 0.08 & 0.03 \\ 
Jackknife & 0.00 & 0.00 & 0.00 & 0.00 & 0.00 & -0.01 & -0.01 & 0.00 & 0.00 & 0.00 & -0.01 & -0.01 & 0.00 & 0.00 & 0.00 & -0.03 & -0.03 & 0.00 & 0.00 & 0.00 & 0.11 & 0.10 & 0.07 & 0.04 & 0.01 & 0.20 & 0.18 & 0.12 & 0.08 & 0.03 \\ 
Bootstrap & 0.00 & 0.00 & 0.00 & 0.00 & 0.00 & -0.01 & -0.01 & 0.00 & 0.00 & 0.00 & -0.01 & -0.01 & 0.00 & 0.00 & 0.00 & -0.03 & -0.03 & -0.01 & 0.00 & 0.00 & 0.11 & 0.10 & 0.07 & 0.05 & 0.01 & 0.20 & 0.18 & 0.12 & 0.08 & 0.03 \\ 
Ozenne et al. & 0.00 & 0.00 & 0.00 & 0.00 & 0.00 & -0.01 & -0.01 & 0.00 & 0.00 & 0.00 & -0.01 & -0.01 & 0.00 & 0.00 & 0.00 & -0.03 & -0.03 & 0.00 & 0.00 & 0.00 & 0.11 & 0.10 & 0.07 & 0.04 & 0.01 & 0.20 & 0.18 & 0.12 & 0.08 & 0.03 \\ 
REML & 0.00 & 0.00 & 0.00 & 0.00 & 0.00 & -0.01 & -0.01 & 0.00 & 0.00 & 0.00 & -0.01 & -0.01 & 0.00 & 0.00 & 0.00 & -0.03 & -0.03 & 0.00 & 0.00 & 0.00 & 0.11 & 0.10 & 0.07 & 0.04 & 0.01 & 0.20 & 0.18 & 0.12 & 0.08 & 0.03 \\ 
\midrule\addlinespace[2.5pt]
\multicolumn{31}{l}{\(\psi_{11}\)} \\[2.5pt] 
\midrule\addlinespace[2.5pt]
ML & -0.13 & -0.09 & -0.04 & -0.02 & 0.00 & -0.26 & -0.16 & -0.06 & -0.03 & 0.00 & -0.16 & -0.12 & -0.05 & -0.02 & 0.00 & -0.32 & -0.22 & -0.09 & -0.04 & 0.00 & 0.42 & 0.37 & 0.23 & 0.16 & 0.05 & 0.53 & 0.50 & 0.35 & 0.26 & 0.09 \\ 
eRBM & -0.07 & -0.03 & -0.01 & -0.01 & 0.00 & -0.23 & -0.12 & -0.03 & -0.01 & 0.00 & -0.10 & -0.06 & -0.02 & -0.01 & 0.00 & -0.28 & -0.18 & -0.07 & -0.03 & 0.00 & 0.41 & 0.37 & 0.23 & 0.17 & 0.05 & 0.50 & 0.49 & 0.35 & 0.26 & 0.09 \\ 
iRBM & -0.07 & -0.03 & -0.01 & -0.01 & 0.00 & -0.22 & -0.11 & -0.03 & -0.01 & 0.00 & -0.09 & -0.06 & -0.02 & -0.01 & 0.00 & -0.26 & -0.16 & -0.07 & -0.03 & 0.00 & 0.41 & 0.37 & 0.23 & 0.17 & 0.05 & 0.50 & 0.49 & 0.35 & 0.26 & 0.09 \\ 
Jackknife & -0.06 & -0.02 & -0.01 & -0.01 & 0.00 & -0.27 & -0.18 & -0.06 & -0.01 & 0.00 & -0.09 & -0.06 & -0.02 & -0.01 & 0.00 & -0.31 & -0.23 & -0.08 & -0.03 & 0.00 & 0.42 & 0.37 & 0.23 & 0.17 & 0.05 & 0.50 & 0.46 & 0.33 & 0.26 & 0.09 \\ 
Bootstrap & -0.06 & -0.02 & -0.01 & -0.01 & 0.00 & -0.23 & -0.17 & -0.05 & -0.01 & 0.00 & -0.09 & -0.06 & -0.02 & -0.01 & 0.00 & -0.29 & -0.21 & -0.08 & -0.03 & 0.00 & 0.43 & 0.37 & 0.23 & 0.17 & 0.05 & 0.52 & 0.47 & 0.34 & 0.26 & 0.09 \\ 
Ozenne et al. & -0.06 & -0.02 & -0.01 & -0.01 & 0.00 & -0.21 & -0.11 & -0.03 & -0.01 & 0.00 & -0.09 & -0.06 & -0.02 & -0.01 & 0.00 & -0.27 & -0.18 & -0.07 & -0.03 & 0.00 & 0.41 & 0.37 & 0.23 & 0.17 & 0.05 & 0.51 & 0.49 & 0.35 & 0.26 & 0.09 \\ 
REML & -0.03 & -0.02 & -0.01 & -0.01 & 0.00 & -0.08 & -0.06 & -0.03 & -0.01 & 0.00 & -0.07 & -0.06 & -0.02 & -0.01 & 0.00 & -0.19 & -0.13 & -0.07 & -0.03 & 0.00 & 0.44 & 0.38 & 0.23 & 0.17 & 0.05 & 0.61 & 0.54 & 0.35 & 0.26 & 0.09 \\ 
\midrule\addlinespace[2.5pt]
\multicolumn{31}{l}{\(\psi_{22}\)} \\[2.5pt] 
\midrule\addlinespace[2.5pt]
ML & -0.09 & -0.07 & -0.03 & -0.01 & 0.00 & -0.16 & -0.13 & -0.05 & -0.04 & 0.00 & -0.11 & -0.09 & -0.04 & -0.01 & 0.00 & -0.26 & -0.21 & -0.09 & -0.07 & 0.00 & 0.34 & 0.30 & 0.19 & 0.13 & 0.04 & 0.52 & 0.47 & 0.33 & 0.24 & 0.08 \\ 
eRBM & -0.02 & -0.02 & -0.01 & 0.00 & 0.00 & -0.10 & -0.09 & -0.03 & -0.03 & 0.00 & -0.05 & -0.04 & -0.02 & 0.00 & 0.00 & -0.20 & -0.17 & -0.07 & -0.06 & 0.00 & 0.34 & 0.30 & 0.19 & 0.13 & 0.04 & 0.53 & 0.47 & 0.34 & 0.24 & 0.08 \\ 
iRBM & -0.04 & -0.03 & -0.01 & 0.00 & 0.00 & -0.12 & -0.10 & -0.03 & -0.03 & 0.00 & -0.06 & -0.05 & -0.02 & 0.00 & 0.00 & -0.21 & -0.17 & -0.07 & -0.06 & 0.00 & 0.34 & 0.30 & 0.19 & 0.13 & 0.04 & 0.52 & 0.47 & 0.33 & 0.24 & 0.08 \\ 
Jackknife & -0.02 & -0.01 & -0.01 & 0.00 & 0.00 & -0.10 & -0.08 & -0.03 & -0.03 & 0.00 & -0.04 & -0.04 & -0.02 & 0.00 & 0.00 & -0.20 & -0.17 & -0.07 & -0.06 & 0.00 & 0.35 & 0.30 & 0.19 & 0.13 & 0.04 & 0.53 & 0.47 & 0.34 & 0.24 & 0.08 \\ 
Bootstrap & -0.03 & -0.02 & -0.01 & 0.00 & 0.00 & -0.11 & -0.08 & -0.03 & -0.03 & 0.00 & -0.06 & -0.04 & -0.02 & 0.00 & 0.00 & -0.22 & -0.17 & -0.08 & -0.05 & 0.00 & 0.34 & 0.30 & 0.19 & 0.13 & 0.04 & 0.53 & 0.47 & 0.34 & 0.24 & 0.08 \\ 
Ozenne et al. & -0.02 & -0.01 & -0.01 & 0.00 & 0.00 & -0.09 & -0.08 & -0.03 & -0.03 & 0.00 & -0.04 & -0.04 & -0.02 & 0.00 & 0.00 & -0.20 & -0.17 & -0.07 & -0.06 & 0.00 & 0.35 & 0.30 & 0.19 & 0.13 & 0.04 & 0.53 & 0.47 & 0.34 & 0.24 & 0.08 \\ 
REML & -0.02 & -0.01 & -0.01 & 0.00 & 0.00 & -0.09 & -0.08 & -0.03 & -0.03 & 0.00 & -0.04 & -0.04 & -0.02 & 0.00 & 0.00 & -0.20 & -0.17 & -0.07 & -0.06 & 0.00 & 0.35 & 0.30 & 0.19 & 0.13 & 0.04 & 0.53 & 0.47 & 0.34 & 0.24 & 0.08 \\ 
\midrule\addlinespace[2.5pt]
\multicolumn{31}{l}{\(\psi_{12}\)} \\[2.5pt] 
\midrule\addlinespace[2.5pt]
ML & -0.01 & 0.00 & -0.03 & 0.00 & 0.00 & -0.07 & -0.06 & -0.04 & -0.02 & 0.00 & -0.03 & 0.01 & -0.03 & 0.00 & 0.00 & -0.15 & -0.13 & -0.06 & -0.05 & 0.00 & 1.50 & 1.35 & 0.87 & 0.62 & 0.20 & 1.41 & 1.27 & 0.87 & 0.63 & 0.20 \\ 
eRBM & 0.01 & 0.01 & -0.02 & 0.00 & 0.00 & -0.06 & -0.05 & -0.03 & -0.02 & 0.00 & -0.01 & 0.03 & -0.02 & 0.00 & 0.00 & -0.14 & -0.12 & -0.05 & -0.04 & 0.00 & 1.60 & 1.42 & 0.89 & 0.63 & 0.20 & 1.50 & 1.34 & 0.89 & 0.64 & 0.20 \\ 
iRBM & 0.02 & 0.03 & -0.02 & 0.00 & 0.00 & -0.04 & -0.02 & -0.03 & -0.02 & 0.00 & 0.00 & 0.02 & -0.02 & 0.00 & 0.00 & -0.09 & -0.07 & -0.05 & -0.04 & 0.00 & 1.57 & 1.39 & 0.88 & 0.63 & 0.20 & 1.47 & 1.31 & 0.88 & 0.64 & 0.20 \\ 
Jackknife & 0.01 & 0.01 & -0.02 & 0.00 & 0.00 & -0.06 & -0.05 & -0.03 & -0.02 & 0.00 & -0.02 & 0.03 & -0.02 & 0.00 & 0.00 & -0.14 & -0.12 & -0.05 & -0.04 & 0.00 & 1.60 & 1.42 & 0.89 & 0.63 & 0.20 & 1.50 & 1.34 & 0.89 & 0.64 & 0.20 \\ 
Bootstrap & 0.00 & 0.01 & -0.02 & 0.00 & 0.00 & -0.06 & -0.05 & -0.03 & -0.02 & 0.00 & -0.02 & 0.03 & -0.03 & 0.00 & 0.00 & -0.14 & -0.12 & -0.06 & -0.04 & 0.00 & 1.58 & 1.42 & 0.89 & 0.63 & 0.20 & 1.47 & 1.34 & 0.89 & 0.64 & 0.20 \\ 
Ozenne et al. & 0.01 & 0.01 & -0.02 & 0.00 & 0.00 & -0.06 & -0.05 & -0.03 & -0.02 & 0.00 & -0.02 & 0.03 & -0.02 & 0.00 & 0.00 & -0.14 & -0.12 & -0.05 & -0.04 & 0.00 & 1.60 & 1.42 & 0.89 & 0.63 & 0.20 & 1.50 & 1.34 & 0.89 & 0.64 & 0.20 \\ 
REML & 0.00 & 0.01 & -0.02 & 0.00 & 0.00 & -0.07 & -0.05 & -0.03 & -0.02 & 0.00 & -0.02 & 0.03 & -0.02 & 0.00 & 0.00 & -0.15 & -0.12 & -0.05 & -0.04 & 0.00 & 1.60 & 1.42 & 0.89 & 0.63 & 0.20 & 1.50 & 1.34 & 0.89 & 0.64 & 0.20 \\ 
\bottomrule
\end{tabular*}

\end{sidewaystable}%

\begin{sidewaystable}[htbp]

\tbl{\label{tbl-bias-growth-50}Results (trimmed) growth curve model reliability 0.50.}
\centering

\fontsize{7pt}{8.4pt}\selectfont
\setlength{\tabcolsep}{2pt}  

\begin{tabular*}{\linewidth}{@{\extracolsep{\fill}}l|rrrrrrrrrrrrrrrrrrrrrrrrrrrrrr}
\toprule
 & \multicolumn{10}{c}{Relative mean bias} & \multicolumn{10}{c}{Relative median bias} & \multicolumn{10}{c}{Relative RMSE} \\ 
\cmidrule(lr){2-11} \cmidrule(lr){12-21} \cmidrule(lr){22-31}
 & \multicolumn{5}{c}{Normal data} & \multicolumn{5}{c}{Non-normal data} & \multicolumn{5}{c}{Normal data} & \multicolumn{5}{c}{Non-normal data} & \multicolumn{5}{c}{Normal data} & \multicolumn{5}{c}{Non-normal data} \\ 
\cmidrule(lr){2-6} \cmidrule(lr){7-11} \cmidrule(lr){12-16} \cmidrule(lr){17-21} \cmidrule(lr){22-26} \cmidrule(lr){27-31}
 & 15 & 20 & 50 & 100 & 1000 & 15 & 20 & 50 & 100 & 1000 & 15 & 20 & 50 & 100 & 1000 & 15 & 20 & 50 & 100 & 1000 & 15 & 20 & 50 & 100 & 1000 & 15 & 20 & 50 & 100 & 1000 \\ 
\midrule\addlinespace[2.5pt]
\multicolumn{31}{l}{\(\theta_{11}\)} \\[2.5pt] 
\midrule\addlinespace[2.5pt]
ML & 0.00 & 0.00 & 0.00 & 0.00 & 0.00 & -0.01 & -0.01 & 0.00 & 0.00 & 0.00 & -0.01 & -0.01 & 0.00 & 0.00 & 0.00 & -0.03 & -0.03 & 0.00 & 0.00 & 0.00 & 0.11 & 0.10 & 0.07 & 0.04 & 0.01 & 0.20 & 0.18 & 0.12 & 0.08 & 0.03 \\ 
eRBM & 0.00 & 0.00 & 0.00 & 0.00 & 0.00 & -0.01 & -0.01 & 0.00 & 0.00 & 0.00 & -0.01 & -0.01 & 0.00 & 0.00 & 0.00 & -0.03 & -0.03 & 0.00 & 0.00 & 0.00 & 0.11 & 0.10 & 0.07 & 0.04 & 0.01 & 0.20 & 0.18 & 0.12 & 0.08 & 0.03 \\ 
iRBM & 0.00 & 0.00 & 0.00 & 0.00 & 0.00 & -0.01 & 0.00 & 0.00 & 0.00 & 0.00 & -0.01 & 0.00 & 0.00 & 0.00 & 0.00 & -0.02 & -0.02 & 0.00 & 0.00 & 0.00 & 0.11 & 0.10 & 0.07 & 0.04 & 0.01 & 0.20 & 0.18 & 0.12 & 0.08 & 0.03 \\ 
Jackknife & 0.00 & 0.00 & 0.00 & 0.00 & 0.00 & -0.03 & -0.02 & 0.00 & 0.00 & 0.00 & -0.01 & -0.01 & 0.00 & 0.00 & 0.00 & -0.04 & -0.03 & 0.00 & 0.00 & 0.00 & 0.11 & 0.10 & 0.07 & 0.04 & 0.01 & 0.19 & 0.17 & 0.12 & 0.08 & 0.03 \\ 
Bootstrap & 0.00 & 0.00 & 0.00 & 0.00 & 0.00 & -0.02 & -0.01 & 0.00 & 0.00 & 0.00 & -0.01 & -0.01 & 0.00 & 0.00 & 0.00 & -0.03 & -0.03 & -0.01 & 0.00 & 0.00 & 0.11 & 0.10 & 0.07 & 0.05 & 0.01 & 0.19 & 0.18 & 0.12 & 0.08 & 0.03 \\ 
Ozenne et al. & 0.00 & 0.00 & 0.00 & 0.00 & 0.00 & -0.01 & -0.01 & 0.00 & 0.00 & 0.00 & -0.01 & -0.01 & 0.00 & 0.00 & 0.00 & -0.03 & -0.03 & 0.00 & 0.00 & 0.00 & 0.11 & 0.10 & 0.07 & 0.04 & 0.01 & 0.20 & 0.18 & 0.12 & 0.08 & 0.03 \\ 
REML & -0.01 & -0.01 & 0.00 & 0.00 & 0.00 & -0.02 & -0.01 & 0.00 & 0.00 & 0.00 & -0.01 & -0.01 & 0.00 & 0.00 & 0.00 & -0.03 & -0.03 & 0.00 & 0.00 & 0.00 & 0.11 & 0.10 & 0.06 & 0.04 & 0.01 & 0.20 & 0.18 & 0.12 & 0.08 & 0.03 \\ 
\midrule\addlinespace[2.5pt]
\multicolumn{31}{l}{\(\psi_{11}\)} \\[2.5pt] 
\midrule\addlinespace[2.5pt]
ML & -0.27 & -0.17 & -0.07 & -0.03 & -0.01 & -0.35 & -0.22 & -0.08 & -0.04 & -0.01 & -0.32 & -0.23 & -0.10 & -0.03 & -0.01 & -0.45 & -0.30 & -0.12 & -0.06 & -0.01 & 0.85 & 0.74 & 0.48 & 0.34 & 0.11 & 0.93 & 0.87 & 0.57 & 0.41 & 0.13 \\ 
eRBM & -0.16 & -0.05 & -0.02 & 0.00 & 0.00 & -0.27 & -0.13 & -0.03 & -0.01 & 0.00 & -0.19 & -0.11 & -0.05 & -0.01 & 0.00 & -0.34 & -0.21 & -0.07 & -0.03 & -0.01 & 0.83 & 0.74 & 0.49 & 0.34 & 0.11 & 0.90 & 0.85 & 0.58 & 0.41 & 0.13 \\ 
iRBM & -0.17 & -0.07 & -0.03 & 0.00 & 0.00 & -0.24 & -0.12 & -0.03 & -0.02 & 0.00 & -0.20 & -0.13 & -0.05 & -0.01 & 0.00 & -0.32 & -0.19 & -0.07 & -0.03 & -0.01 & 0.82 & 0.74 & 0.49 & 0.34 & 0.11 & 0.89 & 0.84 & 0.57 & 0.41 & 0.13 \\ 
Jackknife & -0.12 & -0.05 & -0.02 & 0.00 & 0.00 & -0.28 & -0.17 & -0.03 & -0.01 & 0.00 & -0.16 & -0.11 & -0.04 & -0.01 & 0.00 & -0.36 & -0.24 & -0.07 & -0.03 & -0.01 & 0.86 & 0.74 & 0.49 & 0.34 & 0.11 & 0.90 & 0.82 & 0.57 & 0.41 & 0.13 \\ 
Bootstrap & -0.13 & -0.05 & -0.02 & 0.00 & 0.00 & -0.24 & -0.15 & -0.03 & -0.01 & 0.00 & -0.20 & -0.10 & -0.05 & 0.00 & 0.00 & -0.35 & -0.22 & -0.08 & -0.03 & -0.01 & 0.88 & 0.75 & 0.49 & 0.34 & 0.11 & 0.93 & 0.84 & 0.57 & 0.41 & 0.13 \\ 
Ozenne et al. & -0.14 & -0.05 & -0.02 & 0.00 & 0.00 & -0.24 & -0.12 & -0.02 & -0.01 & 0.00 & -0.18 & -0.10 & -0.04 & -0.01 & 0.00 & -0.32 & -0.20 & -0.07 & -0.03 & -0.01 & 0.83 & 0.75 & 0.49 & 0.34 & 0.11 & 0.91 & 0.86 & 0.58 & 0.41 & 0.13 \\ 
REML & 0.02 & 0.01 & -0.02 & 0.00 & 0.00 & 0.00 & 0.01 & -0.02 & -0.01 & 0.00 & -0.11 & -0.07 & -0.04 & -0.01 & 0.00 & -0.16 & -0.13 & -0.07 & -0.03 & -0.01 & 0.75 & 0.67 & 0.47 & 0.34 & 0.11 & 0.84 & 0.78 & 0.55 & 0.41 & 0.13 \\ 
\midrule\addlinespace[2.5pt]
\multicolumn{31}{l}{\(\psi_{22}\)} \\[2.5pt] 
\midrule\addlinespace[2.5pt]
ML & -0.11 & -0.08 & -0.04 & -0.01 & 0.00 & -0.17 & -0.14 & -0.05 & -0.04 & 0.00 & -0.14 & -0.11 & -0.05 & -0.02 & 0.00 & -0.27 & -0.22 & -0.09 & -0.07 & -0.01 & 0.42 & 0.37 & 0.23 & 0.16 & 0.05 & 0.58 & 0.52 & 0.36 & 0.26 & 0.08 \\ 
eRBM & -0.03 & -0.02 & -0.02 & 0.00 & 0.00 & -0.10 & -0.09 & -0.02 & -0.03 & 0.00 & -0.06 & -0.05 & -0.03 & 0.00 & 0.00 & -0.21 & -0.17 & -0.07 & -0.06 & 0.00 & 0.42 & 0.37 & 0.24 & 0.16 & 0.05 & 0.59 & 0.53 & 0.37 & 0.26 & 0.08 \\ 
iRBM & -0.05 & -0.03 & -0.02 & 0.00 & 0.00 & -0.11 & -0.09 & -0.03 & -0.03 & 0.00 & -0.07 & -0.06 & -0.03 & 0.00 & 0.00 & -0.21 & -0.17 & -0.07 & -0.06 & 0.00 & 0.42 & 0.37 & 0.24 & 0.16 & 0.05 & 0.59 & 0.52 & 0.37 & 0.26 & 0.08 \\ 
Jackknife & -0.03 & -0.02 & -0.02 & 0.00 & 0.00 & -0.09 & -0.08 & -0.02 & -0.03 & 0.00 & -0.05 & -0.05 & -0.03 & 0.00 & 0.00 & -0.20 & -0.17 & -0.07 & -0.06 & 0.00 & 0.42 & 0.37 & 0.24 & 0.16 & 0.05 & 0.59 & 0.53 & 0.37 & 0.26 & 0.08 \\ 
Bootstrap & -0.05 & -0.02 & -0.02 & 0.00 & 0.00 & -0.11 & -0.08 & -0.02 & -0.03 & 0.00 & -0.07 & -0.05 & -0.03 & 0.00 & 0.00 & -0.22 & -0.17 & -0.07 & -0.06 & 0.00 & 0.42 & 0.37 & 0.24 & 0.16 & 0.05 & 0.59 & 0.53 & 0.37 & 0.26 & 0.08 \\ 
Ozenne et al. & -0.03 & -0.02 & -0.02 & 0.00 & 0.00 & -0.09 & -0.08 & -0.02 & -0.03 & 0.00 & -0.05 & -0.05 & -0.03 & 0.00 & 0.00 & -0.20 & -0.17 & -0.07 & -0.06 & 0.00 & 0.42 & 0.37 & 0.24 & 0.16 & 0.05 & 0.59 & 0.53 & 0.37 & 0.26 & 0.08 \\ 
REML & -0.01 & -0.01 & -0.01 & 0.00 & 0.00 & -0.07 & -0.06 & -0.02 & -0.03 & 0.00 & -0.04 & -0.04 & -0.03 & 0.00 & 0.00 & -0.18 & -0.15 & -0.07 & -0.06 & 0.00 & 0.42 & 0.37 & 0.23 & 0.16 & 0.05 & 0.57 & 0.52 & 0.36 & 0.26 & 0.08 \\ 
\midrule\addlinespace[2.5pt]
\multicolumn{31}{l}{\(\psi_{12}\)} \\[2.5pt] 
\midrule\addlinespace[2.5pt]
ML & 0.30 & 0.19 & 0.05 & 0.02 & 0.01 & 0.24 & 0.17 & 0.03 & 0.04 & 0.01 & 0.37 & 0.30 & 0.07 & 0.03 & 0.02 & 0.26 & 0.18 & 0.07 & 0.06 & 0.00 & 2.49 & 2.20 & 1.44 & 1.03 & 0.32 & 2.47 & 2.25 & 1.49 & 1.05 & 0.34 \\ 
eRBM & 0.15 & 0.07 & 0.00 & 0.00 & 0.00 & 0.08 & 0.05 & -0.02 & 0.02 & 0.01 & 0.23 & 0.19 & 0.02 & 0.01 & 0.02 & 0.12 & 0.07 & 0.02 & 0.04 & 0.00 & 2.63 & 2.30 & 1.47 & 1.04 & 0.32 & 2.61 & 2.35 & 1.52 & 1.06 & 0.34 \\ 
iRBM & 0.17 & 0.09 & 0.01 & 0.00 & 0.00 & 0.09 & 0.05 & -0.02 & 0.02 & 0.01 & 0.25 & 0.21 & 0.02 & 0.01 & 0.02 & 0.10 & 0.06 & 0.01 & 0.04 & 0.00 & 2.61 & 2.28 & 1.47 & 1.04 & 0.32 & 2.59 & 2.33 & 1.52 & 1.06 & 0.34 \\ 
Jackknife & 0.14 & 0.07 & 0.00 & -0.01 & 0.00 & 0.07 & 0.05 & -0.02 & 0.02 & 0.01 & 0.22 & 0.18 & 0.02 & 0.01 & 0.02 & 0.12 & 0.07 & 0.01 & 0.04 & 0.00 & 2.64 & 2.30 & 1.47 & 1.04 & 0.32 & 2.63 & 2.35 & 1.52 & 1.06 & 0.34 \\ 
Bootstrap & 0.19 & 0.07 & 0.00 & 0.00 & 0.00 & 0.12 & 0.06 & -0.02 & 0.02 & 0.01 & 0.29 & 0.21 & 0.02 & 0.01 & 0.01 & 0.15 & 0.09 & 0.02 & 0.03 & 0.00 & 2.60 & 2.30 & 1.47 & 1.04 & 0.32 & 2.58 & 2.35 & 1.52 & 1.06 & 0.34 \\ 
Ozenne et al. & 0.14 & 0.07 & 0.00 & -0.01 & 0.00 & 0.07 & 0.05 & -0.02 & 0.02 & 0.01 & 0.22 & 0.18 & 0.02 & 0.01 & 0.02 & 0.12 & 0.07 & 0.01 & 0.04 & 0.00 & 2.64 & 2.30 & 1.47 & 1.04 & 0.32 & 2.63 & 2.35 & 1.52 & 1.06 & 0.34 \\ 
REML & -0.13 & -0.08 & -0.03 & -0.01 & 0.00 & -0.29 & -0.21 & -0.08 & 0.01 & 0.01 & 0.07 & 0.06 & 0.01 & 0.00 & 0.02 & -0.23 & -0.13 & -0.02 & 0.04 & 0.00 & 2.42 & 2.17 & 1.44 & 1.04 & 0.32 & 2.38 & 2.18 & 1.48 & 1.06 & 0.34 \\ 
\bottomrule
\end{tabular*}

\end{sidewaystable}%